\documentclass[prx,twocolumn,superscriptaddress,citeautoscript,amsart,longbibliography]{revtex4-2}

\usepackage[labelfont=bf,font=small]{caption}
\usepackage{graphicx}
\usepackage{multirow}
\usepackage{color}
\usepackage{bm}
\usepackage{times}
\usepackage{amsmath,bm,amsfonts}
\usepackage{dcolumn}
\usepackage{graphicx}
\usepackage{latexsym}
\usepackage{hhline}
\usepackage{braket}
\usepackage{bbold}
\usepackage{multirow}

\def\ket#1{|#1\rangle }
\def\bra#1{\langle #1 |}

\def\bb{\mathbb}

\def\d{\partial}

\renewcommand{\thetable}{\arabic{table}}

\begin{document}
\title{
Riemannian geometry of resonant optical responses
}
\author{Junyeong \surname{Ahn}}
\email{junyeongahn@fas.harvard.edu}
\affiliation{Department of Physics, Harvard University, Cambridge, MA 02138, USA}

\author{Guang-Yu \surname{Guo}}
\email{gyguo@phys.ntu.edu.tw}
\affiliation{Department of Physics and Center for Theoretical Physics, National Taiwan University, Taipei 10617, Taiwan}
\affiliation{Physics Division, National Center for Theoretical Sciences, Taipei 10617, Taiwan}

\author{Naoto \surname{Nagaosa}}
\email{nagaosa@riken.jp}
\affiliation{RIKEN Center for Emergent Matter Science (CEMS), Wako, Saitama 351-0198, Japan}
\affiliation{Department of Applied Physics, The University of Tokyo, Bunkyo, Tokyo 113-8656, Japan}

\author{Ashvin \surname{Vishwanath}}
\email{avishwanath@g.harvard.edu}
\affiliation{Department of Physics, Harvard University, Cambridge, MA 02138, USA}

\date{\today}

\begin{abstract}
{\bf The geometry of quantum states is well-established as a basis for understanding the response of electronic systems to static electromagnetic fields, as exemplified by the theory of the quantum and anomalous Hall effects. However, it has been challenging to relate quantum geometry to resonant optical responses. The main obstacle is that optical transitions involve a pair of states, while existing geometrical properties are defined for a single state. As a result, a concrete geometric understanding of optical responses has so far been limited to two-level systems, where the Hilbert space is completely determined by a single state and its orthogonal complement. Here, we construct a general theory of Riemannian geometry for resonant optical processes by identifying transition dipole moment matrix elements as tangent vectors. This theory applies to arbitrarily high-order responses, suggesting that optical responses can generally be thought of as manifestations of the Riemannian geometry of quantum states. We use our theory to show that third-order photovoltaic Hall effects are related to the Riemann curvature tensor and demonstrate an experimentally accessible regime where they dominate the response.}
\end{abstract}

\maketitle

A common feature of geometric responses~\cite{thouless1982quantized,nagaosa2010anomalous,xiao2010berry,sodemann2015quantum,
hasan2010colloquium,armitage2018weyl,
neupert2013measuring,
peotta2015superfluidity,xie2020topology,
lapa2019semiclassical,gao2019nonreciprocal,zhao2020electric,kozii2020intrinsic,
gao2015geometrical,rhim2020quantum} is that they originate from inter-band hybridizations, such that they appear as quantum corrections to semiclassical descriptions.
Since resonant optical responses are inherently inter-band quantum-mechanical processes, they are ideal candidates for a geometrical interpretation.
Indeed, nonlinear optical responses, in particular, have attracted much attention recently for this reason~\cite{hosur2011circular,morimoto2016topological,nagaosa2017concept,ahn2020low,de2017quantized,de2020difference,flicker2018chiral,holder2020consequences,watanabe2021chiral} in addition to the possibility of diverse physical applications~\cite{sturman1992photovoltaic,tokura2018nonreciprocal,boyd2020nonlinear}.
Quantized injection photocurrent responses were found in chiral semimetals~\cite{de2017quantized,flicker2018chiral,de2020difference}, and the shift photovoltaic responses of semimetals have been found to be described by the Levi-Civita connection~\cite{ahn2020low}, a geometric quantity distinguished from the typical Berry curvature and quantum metric.

However, the study of the relationship between quantum geometry and optical responses has been limited to the vicinity of crossing points in semimetals, where two-band descriptions are good approximations~\cite{de2017quantized,de2020difference,ahn2020low} or special selection rules exist due to linear dispersion relations~\cite{flicker2018chiral}.
The fundamental reason for this is the absence of a proper theory explaining the geometry of optical transitions; the geometric quantities like the quantum metric and the Berry curvature are usually defined for a single quantum state~\cite{provost1980riemannian}, while optical transitions involve a pair of states.
While the wave function of one state completely determines the other in two-band systems, in general the two quantum states involved in an optical transition are independent, meaning that the geometry of a pair of states is different from the geometry of a single state.
Additional constraints like the strictly linear dispersion relations was required for a geometric interpretation of optical responses in more-than-two-band systems such as chiral multifold fermions~\cite{flicker2018chiral}.

\begin{figure*}[t!]
\includegraphics[width=\textwidth]{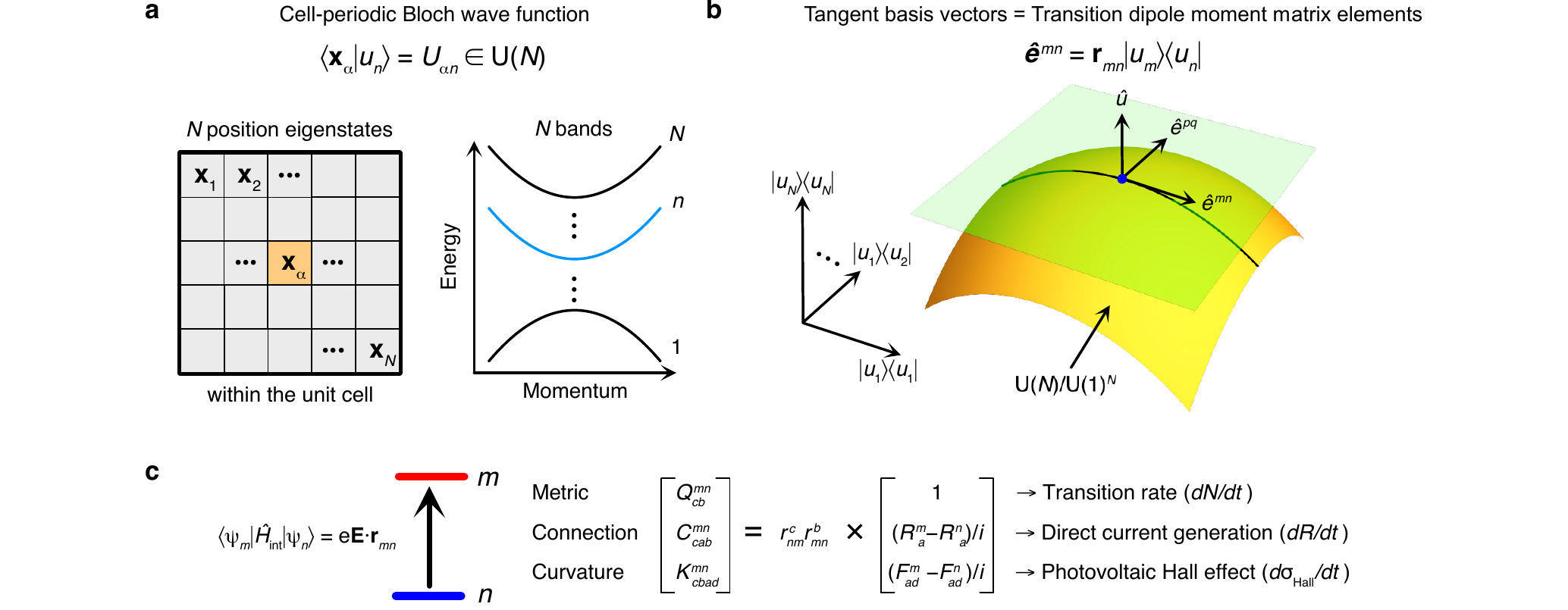}
\caption{
{\bf Geometry of the cell-periodic Bloch state and optical transitions.}
{\bf a}, Cell-periodic Bloch wave function as an $N$ by $N$ unitary matrix.
The limit $N\rightarrow \infty$ is assumed.
{\bf b}, Transition dipole moment matrix elements as tangent basis vectors.
$\hat{e}^{pq}$s and $\hat{e}^{mn}$ are tangent vectors on the manifold ${\rm U}(N)/{\rm U}(1)^N$ of cell-periodic Bloch states, and $\hat{u}$ presents normal vectors.
Here, ${\rm U}(1)^N$ corresponds to the gauge degrees of freedom.
Optical transitions between $m$ and $n$ states reveal geometrical information on the curve generated by $\hat{e}^{mn}$.
{\bf c}, Physical meaning of the Hermitian metric, connection, and curvature in optical processes for non-degenerate states.
$R_a^{n}$ and $F^n_{ad}$ are the Wannier center and the Berry curvature of the state $\ket{u_n}$.
}
\label{fig:geometry}
\end{figure*}

In this Article, we introduce a general theoretical description of the Riemannian geometry for resonant optical transitions.
We consider independent electronic quasiparticles at zero-temperature in clean crystals and neglect photon momentum.
Let us note that optical responses are described by the electric dipole Hamiltonian~\cite{aversa1995nonlinear,ventura2017gauge}
\begin{align}
\hat{H}_{\rm int}=e{\bf E}(t)\cdot\hat{\bf r},
\end{align}
where $-e$ is the electric charge (Methods 1 summarizes our conventions and definitions), and $\hat{\bf r}$ is the position operator of electrons, and ${\bf E}(t)$ is the external electric field.
Since all information on the unperturbed quantum states is encoded in the overlap matrices $\braket{\psi_{m\bf k}|H_{\rm int}|\psi_{n\bf k}}$ in perturbative response theory, the transition dipole moment matrix element (per unit charge)
\begin{align}
{\bf r}_{mn}({\bf k})
=\braket{\psi_{m\bf k}|\hat{\bf r}|\psi_{n\bf k}}
\end{align}
is the fundamental building block of the $k$th-order nonlinear optical conductivity tensors~\cite{boyd2020nonlinear} defined by (see Methods 2)
\begin{align}
j^{a}(\omega)
&=
\sum_k
\sum_{a_i,\omega_i}
\sigma^{a;a_1,\hdots,a_k}(\omega;\omega_1,\hdots,\omega_k)
E^{a_1}(\omega_1)\hdots E^{a_k}(\omega_k).
\end{align}

Our key observation is that the transition dipole moment matrix elements define tangent vectors on the manifold of quantum states.
In differential geometry, the tangent vector associated with a coordinate $k^a$ is written in an abstract way as $\frac{\d}{\d k_a}$ for $a=1,\hdots,d$, where $d$ is the spatial dimension.
Therefore, one can expect that the position operator $\hat{r}^a\sim  i\d_a\equiv i\frac{\d}{\d k_a}$ defines the tangent vector for the momentum coordinate $k^a$ defined on a manifold.
Then, defining a metric tensor defines the Riemannian structure.
We make this idea concrete for the representation of $\hat{r}^a$ on the cell-periodic Bloch states.
The same procedure may also be applied to the energy eigenstates in a ${\bf k \cdot p}$ model or a tight-binding model.

This construction allows us to show that resonant optical responses are geometric in nature.
We show that linear optical responses and second-order bulk photovoltaic effects are manifestations of the metric and connection.
In particular, the non-quantized injection photocurrents in chiral semimetals due to higher-band corrections to the two-band description~\cite{de2017quantized,de2020difference} or a quadratic dispersion in chiral multifold fermion models~\cite{flicker2018chiral} can still be regarded as geometric responses, although not as topological responses.
Furthermore, our geometric perspective can serve as a useful organizing principle for more complicated higher-order optical conductivity tensors.
As an example, we show that Riemann curvature characterizes the third-order photovoltaic Hall conductivity near the band edge of gapped topological materials.

To simplify the presentation of our theory, we mainly consider non-degenerate states without spacetime inversion $(PT)$ symmetry in the main text.
See Methods 5 and 6 for details on degenerate states and symmetry constraints.
\\

{\bf \noindent Transition dipole moment as a tangent vector.}
We begin by revisiting the geometric meaning of the position operator.
The position operator is represented in the Bloch state basis as~\cite{karplus1954hall,blount1962formalisms}
\begin{align}
r^a_{mn}({\bf k})=\delta_{mn}i\d_a+\braket{u_{m\bf k}|i\d_a|u_{n\bf k}}.
\end{align}
The second term is a geometric quantity widely known as the non-abelian Berry connection.
Here, instead of viewing it as a connection, we regard it as the Maurer-Cartan form $\xi_a({\bf k})=U^{\dagger}({\bf k})i\d_aU({\bf k})$ (see Methods 4) of the $N\times N$ unitary matrix $U_{\alpha m}({\bf k})=\braket{{\bf x}_{\alpha}|u_{m\bf k}}$, where $\ket{{\bf x}_{\alpha}}$ is a position eigenstate within the unit cell with eigenvalue ${\bf x}_{\alpha}$.
Here, we take both the number of bands and the number of position eigenstates in the unit cell to be a finite number $N$ for definiteness [Fig.~1(a)].
The limit $N\rightarrow \infty$ is taken in the end.
The Maurer-Cartan form has one-to-one correspondence with the tangent vector $i\d_aU({\bf k})$ at $U({\bf k})$, so it can be simply regarded as a tangent vector.
After modding out the ${\rm U}(1)$ gauge degrees of freedom of each state from ${\rm U}(N)$, we obtain the manifold ${\cal M}={\rm U}(N)/{\rm U}(1)^N$ of the cell-periodic Bloch states.
The tangent vectors in ${\cal M}$ are given $\xi^{\cal M}_a({\bf k})=U^{\dagger}({\bf k})i\d_aU({\bf k})-U^{\dagger}({\bf k})i\d_aU({\bf k})|_{{\rm U}(1)^N}$, where $[U^{\dagger}({\bf k})i\d_aU({\bf k})|_{{\rm U}(1)^N}]_{mn}=\delta_{mn}\braket{u_{m\bf k}|i\d_a|u_{n\bf k}}$ is the projection of $\xi_a$ along the ${\rm U}(1)^N$ direction.
In the operator form, the tangent vector is $\hat{\xi}^{\cal M}_a({\bf k})=\sum_{m\ne n}r^a_{mn}({\bf k})\ket{u_{m\bf k}}\bra{u_{n\bf k}}$.

Now we arrive at our main result: the transition dipole moment $r^a_{mn}$ between $m$ and $n$ states is one particular component of the $N(N-1)/2$-dimensional complex-valued tangent vector $\hat{\xi}_a^{\cal M}$ (i.e., $r^a_{mn}$ is a complex-valued vielbein).
Namely, an optical transition between $m$ and $n$ states probes the one-dimensional complex vector space spanned by
\begin{align}
\hat{e}^{mn}_a({\bf k})\equiv r^a_{mn}({\bf k})\ket{u_{m\bf k}}\bra{u_{n\bf k}}, \quad a=1,\hdots,d,
\end{align}
with given $m$ and $n$ with $m\ne n$ [Fig.~1(b)].
In the following, we describe the Riemannian structure of the subspace spanned by $\hat{e}^{mn}_a$s.
Here, at most two $\hat{e}^{mn}_{a=1,\hdots,d}$s are linearly independent for a given pair $(m,n)$ because one complex dimension is fully covered by two independent real coordinates.\\

{\bf \noindent Complex Riemannian structure.}
The complex Riemannian structure is induced by the natural inner product in the space of $N\times N$ matrices, called the Hilbert-Schmidt inner product:
\begin{align}
(A,B)
=\sum_{\alpha,\beta}A^*_{\alpha\beta}B_{\alpha\beta}
={\rm Tr}\left[A^{\dagger}B\right].
\end{align}
It is the complex Euclidean inner product with $\ket{{\bf x}_{\alpha}}\bra{{\bf x}_{\beta}}$s as basis vectors (equivalently, $\ket{u_{p\bf k}}\bra{u_{q\bf k}}$s at any fixed ${\bf k}$ as basis vectors).
This inner product is Hermitian as it satisfies $(B,A)^*=(A,B)$.
The Hermitian metric tensor (a.k.a. quantum geometric tensor) in the tangent subspace spanned by $\hat{e}^{mn}_a$s is defined by Hilbert-Schmidt inner product of the tangent basis vectors:
\begin{align}
\label{eq:optical-metric}
Q^{mn}_{ba}
\equiv (\hat{e}^{mn}_b,\hat{e}^{mn}_a)
=r^b_{nm}r^a_{mn}.
\end{align}
This metric is identical to the Fubini-Study metric of the state $\ket{u_{n}}$ in two-band systems, but the two are generally different; the Fubini-Study metric is given by summing $Q^{mn}_{ba}$ over all possible intermediate states $\ket{u_m}$.

The covariant derivative of $\hat{e}^{mn}_a$s define other geometric quantities such as the connection and curvature.
By taking a derivative on $\hat{e}^{mn}_a$, one obtains parallel and perpendicular components:
$\d_c\hat{e}^{mn}_a=\sum_b(C^{mn})^b_{ca}\hat{e}^{mn}_b+\hdots$ where the ellipsis indicates the components perpendicular to $\hat{e}^{mn}_b$s. [Fig.~1].
Then, the covariant derivative $\nabla$ is defined as the parallel-transported part by $\nabla_c\hat{e}^{mn}_a=\sum_b(C^{mn})^b_{ca}e_b$.
The Hermitian connection is defined by
\begin{align}
C^{mn}_{bca}\equiv 
\sum_eQ^{mn}_{be}(C^{mn})^e_{ca}
=(\hat{e}^{mn}_b,\nabla_c\hat{e}^{mn}_a)
=r^b_{nm}r^a_{mn,c},
\end{align}
where $O_{mn,c}\equiv \d_cO_{mn}-i[{\cal A}_c,O]_{mn}$ is called the generalized derivative~\cite{aversa1995nonlinear}, where $({\cal A}_c)_{mn}=\delta_{mn}\braket{u_m|i\d_c|u_n}$ is the ${\rm U}(1)^N$ Berry connection.
Our construction gives a definite geometric meaning to the generalized derivative.
This Hermitian connection is in general different from the Levi-Civita connection, the unique torsionless connection determined by the metric, because it has a nontrivial torsion $T^{mn}_{bca}=C^{mn}_{bca}-C^{mn}_{bac}=ir^b_{nm}\sum_{p\ne m,n}\left(r^c_{mp}r^a_{pn}-r^a_{mp}r^c_{pn}\right)$ originating from virtual transitions among three states.
The Hermitian curvature tensor is defined by antisymmetrizing the second-order covariant derivatives.
\begin{align}
K^{mn}_{badc}
\equiv (\hat{e}^{mn}_b,(\nabla_d\nabla_c-\nabla_c\nabla_d)\hat{e}^{mn}_a)
=-ir^{b}_{nm}[{\cal F}_{dc},r^{a}]_{mn},
\end{align}
where $({\cal F}_{dc})_{pq}=\d_d({\cal A}_c)_{pq}-\d_c({\cal A}_d)_{pq}$ is the ${\rm U}(1)^N$ Berry curvature.
\\

{\bf \noindent Meaning of metric, connection, and curvature in optical processes.}
$Q$, $C$, and $K$ defined here are properties of a one-dimensional complex vector space.
Their real part define the Riemannian metric tensor (a.k.a quantum metric tensor), metric connection, and the Riemann curvature tensor of the corresponding two-dimensional real vector space spanned by ${\rm Re}[\hat{e}_a]$s and ${\rm Im}[\hat{e}_a]$s.
On the other hand, the (minus) imaginary parts define the symplectic form, almost symplectic connection, and the symplectic curvature tensor~\cite{bieliavsky2006symplectic}.
All are gauge invariant.

\begin{figure*}[t!]
\includegraphics[width=\textwidth]{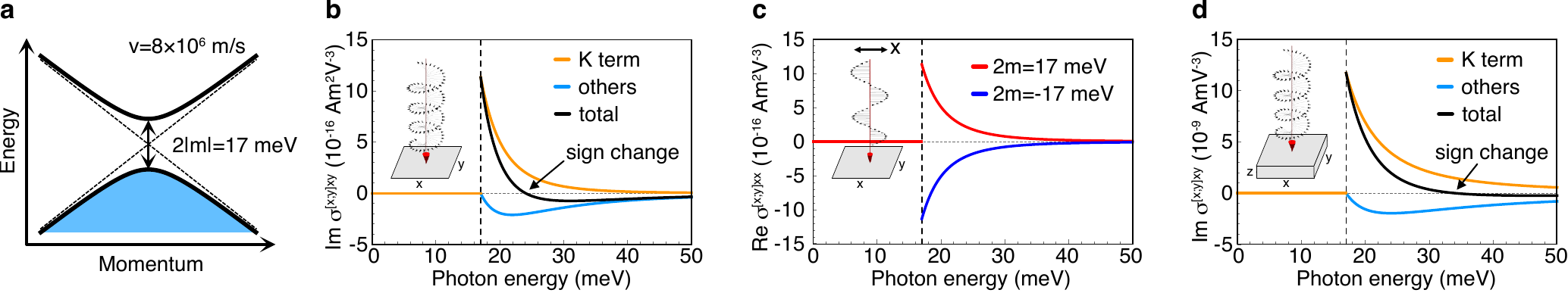}
\caption{
{\bf Third-order photovoltaic Hall conductivity of a Dirac fermion.}
{\bf a}, Band structure of a massive Dirac fermion.
The blue shaded region shows the occupied states.
We take the asymptotic velocity $v=8\times 10^6\;{\rm m/s}$ and the band gap $2|m|=17\;{\rm meV}$ for both 2D and 3D Dirac fermions.
{\bf b} and {\bf c}, Two dimensional Dirac fermion.
Incident light is circularly polarized in {\bf b} and linearly polarized in {\bf c}.
{\bf d}, Three dimensional Dirac fermion.
The conductivity of a single Weyl fermion is the half of that in {\bf d}.
We take the relaxation rate of $\hbar \Gamma=1\;{\rm meV}$ in all calculations.
All non-vanishing tensor components not shown in the figures are related to the shown components by rotational symmetries.}
\label{fig:PVHE}
\end{figure*}

As the geometric quantities defined above are given by the transition dipole moment and its gauge-invariant derivatives, they are basic building blocks of the optical conductivity tensors.
The Hermitian metric appears in the linear optical conductivity tensor as
\begin{align}
\sigma^{b;a}
=\frac{\pi \omega e^2}{h}\sum_{m,n}\int_{\bf k} \delta\left(\omega-\omega_{mn}\right)f_{nm}Q^{mn}_{ba},
\end{align}
where $\int_{\bf k}=\int d^dk/(2\pi)^d$, $\hbar\omega_{mn}$ is the energy difference between $m$ and $n$ bands, and $f_{nm}=f_n-f_m$ is the difference between the Fermi-Dirac distribution of the $n$ and $m$ states.

Also, the Hermitian metric is directly related to the optical transition rate by the Fermi's Golden rule since $|\braket{\psi_{m\bf k}|e\hat{\bf r}\cdot {\bf E}(\omega)|\psi_{n\bf k}}|^2\propto Q^{mn}_{ba}E_b(\omega)E_a^*(\omega)$.
This relation is particularly useful for the interpretation of geometric quantities [Fig.~1(c)].
For example, let us consider the injection photovoltaic effect, the photocurrent generation by the optical transition of the electron velocity.
The corresponding second-order conductivity tensor can be expressed in terms of the Hermitian metric as
\begin{align}
\sigma^{c;ab}_{\rm inj}=-\frac{\pi e^3}{\hbar^2\Gamma}\sum_{m,n}\int_{\bf k} \delta\left(\omega-\omega_{mn}\right)f_{nm}Q^{mn}_{ba}(v^c_{mm}-v^c_{nn}),
\end{align}
where $v^c_{mm}$ and $v^c_{nn}$ are the group velocities of bands $m$ and $n$, respectively, which involve the band dispersion.

To explain the Hermitian connection, we write it as $C_{bca}=r^b_{nm}r^a_{mn,c}=-ir^b_{nm}r^a_{mn}R^{c,a}_{mn}$, where $R^{c,a}_{mn}={\cal A}^c_{mm}-{\cal A}^c_{nn}+i\d_c\log r^a_{mn}$ is the shift vector describing the shift of the electron position during the optical excitation (note that $R^{c,a}_{mn}\sim r^c_{mm}-r^c_{nn}$)~\cite{sturman1992photovoltaic,sipe2000second}.
As $r^b_{nm}r^a_{mn}$ is proportional to the transition rate, $C_{bca}$ is responsible for the shift photovoltaic effect, a generation of the direct current $j^c_{\rm shift}\propto \d R^c/dt$ by illumination of light:
\begin{align}
\sigma^{c;ab}_{\rm shift}
=-\frac{\pi e^3}{2\hbar^2}\sum_{m,n}\int_{\bf k} \delta\left(\omega-\omega_{mn}\right)f_{nm}i (C^{mn}_{bca}-(C^{mn}_{acb})^*).
\end{align}

While the above relations generalize the ones found in two-band systems~\cite{de2017quantized,de2020difference,ahn2020low}, the optical manifestation of the Hermitian curvature has not been known even in two-band systems.
However, we can also interpret the Hermitian curvature
$K_{badc}=-ir^{b}_{nm}[{\cal F}_{dc},r^{a}]_{mn}=-ir^{b}_{nm}r^{a}_{mn}({\cal F}^{dc}_{mm}-{\cal F}^{dc}_{nn})$ in the same vein as the optical transition of the Berry curvature [Fig.~1(c)].
Since the Berry curvature is the source of the Hall effect, the Hermitian curvature is expected to be responsible for the light-induced dc Hall effect, also called the photovoltaic Hall effect~\cite{oka2009photovoltaic}, which grows linearly in time with constant light intensity.
At finite relaxation rate $\Gamma$ of electronic quasiparticles, the saturated photovoltaic Hall conductivity is proportional to $\Gamma^{-1}$; we call this as the injection photovoltaic Hall conductivity following Ref.~\cite{fregoso2019bulk}.
We elaborate more on this response below.\\

{\bf \noindent Hermitian curvature in the photovoltaic Hall effect.}
The expression of the third-order injection conductivity tensor in time-reversal-symmetric systems was derived in Ref.~\cite{fregoso2019bulk}.
When generalized to include time-reversal-breaking systems, the third-order injection conductivity tensor contains the Hermitian curvature.
\begin{align}
\label{eq:third-order-photoconductivity-main}
\sigma^{d;abc}_{\rm inj}
&=
\frac{\pi e^4}{6\Gamma\hbar^3}\sum_{m,n}\int_{\bf k}\delta\left(\omega-\omega_{mn}\right)f_{nm}
iK^{mn}_{cbad}
+\hdots,
\end{align}
where the index $a$ is for the static electric field, indices $b$ and $c$ are for the oscillating electric field of light, the ellipsis includes the second-order connection, connection, metric, and virtual transitions among three states (see Methods 3).
The real (imaginary) part of the tensor is responsible for the response independent (dependent) on the light helicity, which we call linear (circular) photoconductivity.
The photovoltaic Hall response is characterized by the anti-symmetric part $\sigma^{[d;a]bc}_{\rm inj}=(\sigma^{d;abc}_{\rm inj}-\sigma^{a;dbc}_{\rm inj})/2$.
In the clean limit where $\hbar\Gamma$ is much smaller than the photon energy and band gaps, injection response is the largest contribution to the photovoltaic Hall response.

\begin{table}[b!]
\begin{tabular}{c|c|cc|cc}
\multirow{2}*{Response}		& Jerk
&\multicolumn{2}{c|}{Injection}	&\multicolumn{2}{c}{Shift}\\
&Linear
&Linear
&Circular
&Linear
&Circular\\
\hhline{=|=|==|==}
Photovoltaic Hall effect
&No
&\multicolumn{4}{c}{Yes}\\
\hline
$T$ or $PT$ symmetry
&Yes
&No
&Yes
&Yes
&No
\end{tabular}
\caption{
{\bf Properties of the third-order photoconductivity tensors.}
Following Ref.~\cite{fregoso2019bulk}, the third-order photoconductivity is classified into jerk, injection, and shift according to their dependence on the relaxation rate; they are proportional to $\Gamma^{-2}$, $\Gamma^{-1}$, and $\Gamma^0$, respectively.
Linear (circular) means the light-helicity-independent (-dependent) response.
$T$ and $P$ indicate time reversal and spatial inversion.
Since third-order optical conductivity tensors are invariant under spatial inversion, time-reversal-symmetric responses are spacetime-inversion-symmetric also.}
\label{tab:T-symmetry}
\end{table}

In time-reversal-symmetric systems, the injection response depends on the helicity of the circularly polarized light [Table.~\ref{tab:T-symmetry}].
Similarly, spacetime-inversion symmetry also allows only circular photoconductivity because the third-order optical conductivity tensor is invariant under spatial inversion.
On the other hand, when time reversal symmetry and spacetime inversion symmetry are both broken, linearly polarized light can also induce injection photovoltaic Hall conductivity.

\begin{figure*}[t!]
\includegraphics[width=\textwidth]{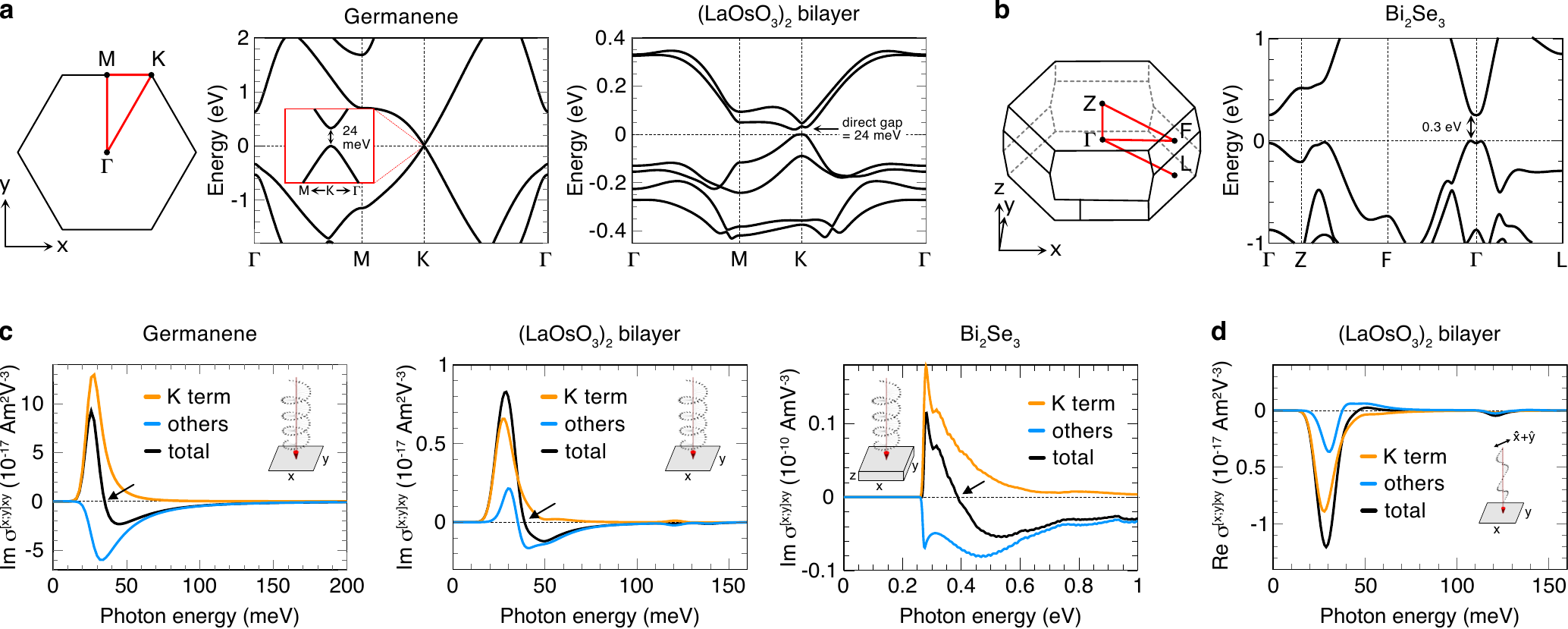}
\caption{
{\bf First-principles calculations on massive Dirac materials.}
Band structures of ({\bf a}) monolayer germanene and ferromagnetic (LaOsO$_3$)$_2$ bilayer and ({\bf b}) bulk Bi$_2$Se$_3$ along the high-symmetry lines.
The top of valence bands is set to $0 \;{\rm eV}$.
{\bf c}, Third-order circular photovoltaic Hall conductivity tensors.
Black arrows point to the sign change of the third-order conductivity tensor.
{\bf d}, Third-order linear photovoltaic Hall conductivity tensor.
Only (LaOsO$_3$)$_2$ bilayer shows nonzero response.
The Gaussian broadening of $5\; {\rm meV}$ is introduced to numerically calculate the the delta function.
We take the relaxation energy scale to be $\hbar\Gamma = 1{\rm \; meV}$ in all cases.
}
\label{fig:ab-initio}
\end{figure*}

Equation~\eqref{eq:third-order-photoconductivity-main} in general has a complicated form including various geometric quantities.
However, as we show now, the Hermitian curvature dominates the photovoltaic Hall response near the band edge of topological materials such as topological insulators and massive Dirac semimetals, characterized by massive Dirac Hamiltonians.

Let us consider the model of a two-dimensional Dirac fermion [Fig.~2(a)].
\begin{align}
\label{eq:2DDirac}
H({\bf k})=\hbar v (k_x\sigma_x+k_y\sigma_y)+m\sigma_z.
\end{align}
As the nonzero mass term break time reversal $T=i\sigma_yK$ and spacetime inversion $PT=\sigma_xK$ symmetries, both linear and circular photovoltaic Hall effects can occur.

We first consider circularly polarized light.
Figure~2(b) shows the third-order photovoltaic Hall conductivity tensors calculated with $\hbar\Gamma=1\;{\rm meV}$, $v=8\times 10^5\;{\rm m/s}$, and $2|m|=17\;{\rm meV}$, which are relevant to graphene on a hexagonal boron nitride substrate~\cite{kim2018accurate}.
The Hermitian curvature dominates the response near the band edge, while the other contributions having the opposite sign grow as the photon energy goes higher.
This leads to the sign change of the third-order conductivity at $2\sqrt{2}|m|=24\;{\rm meV}$.
Since the circular photovoltaic Hall conductivity is $T$- and $PT$-symmetric, it is independent of the sign of the mass $m$.
Note that the response at high photon energies follows Oka and Aoki's result~\cite{oka2009photovoltaic} for massless graphene.
Therefore, the opposite sign of the response at the band edge is an experimentally observable signature that the response is due to the Hermitian curvature and is not by the Oka-Aoki mechanism.

On the other hand, the linear photovoltaic Hall conductivity tensor changes sign as the mass sign changes, reflecting the fact that the linear photovoltaic Hall effect is a time-reversal-breaking effect [Fig.~2(c)].
Another feature of the linear photovoltaic Hall conductivity in Fig.~2(c) is that no sign change occurs as the photon energy increases above the band gap.
This is because the response is purely from the Hermitian curvature in our model.

Three-dimensional massive Dirac fermions show responses similar to two-dimensional Dirac fermions [Fig.~2(d)].
However, their linear photovoltaic Hall conductivity is zero because of spacetime inversion symmetry.
See Methods 8 for analytic expressions of conductivity tensors.

We test our model-based predictions by performing first-principles calculations for monolayer germanene, bulk Bi$_2$Se$_3$, and (LaOsO$_3$)$_2$ bilayer [Fig.~3].
These materials are ${\bb Z}_2$ topological insulators in two and three dimensions~\cite{acun2015germanene,xia2009observation}, and a Chern insulator~\cite{chandra2017quantum}, respectively, described by a massive Dirac Hamiltonian close to the band edge.
As expected from model calculations, the Hermitian curvature dominates the response near the band edge, and the third-order circular photovoltaic Hall conductivity changes sign as the photon energy increases [Fig.~3(c)].
Ferromagnetic (LaOsO$_3$)$_2$ bilayer can additionally show the linear photovoltaic Hall effect because time reversal symmetry is broken [Fig.~3(d)].
Our calculations demonstrate that the Hermitian curvature is a useful measure of the photovoltaic Hall response in topological materials.\\

{\bf \noindent Topological optical response.}
Given an understanding of geometry, one interesting question is about its topology.
Topological invariants can be defined using standard methods in Riemannian geometry.
Recall the target geometry probed by the transition between a specific pair of states is complex one dimension.
The Euler number is then a natural topological invariant to consider because it classifies all one-dimensional closed complex manifolds (but note that we probe a one-dimensional complex subtangent space, which may not be the tangent space of a one-dimensional complex manifold. See Supplementary Note 1 for more discussions.).
According to the Gauss-Bonnet theorem, the Euler number for the transition between $n$ and $m$ states is given by
\begin{align}
\label{eq:Gauss-Bonnet}
\chi^{mn}
=\frac{1}{2\pi}\oint dk_1\wedge dk_2\frac{R^{mn}_{1212}}{\sqrt{g}},
\end{align}
where $k_1$ and $k_2$ are two independent coordinates parametrizing a two-dimensional submanifold in momentum space, $dk_1\wedge dk_2=dk_1dk_2{\rm sgn}(-{\rm Im}Q^{mn}_{12})$ is the oriented area where the sign determines the orientation of the mapping from momentum space to the manifold of quantum states, $g\equiv \det g^{mn}=g^{mn}_{11}g^{mn}_{22}-(g^{mn}_{12})^2$, $g^{mn}_{ij}={\rm Re}Q^{mn}_{ij}$ is the Riemannian metric, and
$R^{mn}_{1212}={\rm Re}K^{mn}_{1212}$ is the Riemann curvature tensor.
This invariant is relevant only in three dimensions because optical transitions in $d$ spatial dimensions occurs over $(d-1)$-dimensional submanifolds in momentum space.

We note that the integrand in Eq.~\eqref{eq:Gauss-Bonnet} cannot be written as a polynomial function of geometric quantities defined for the transition between $n$ and $m$ states.
This property implies that the topological optical responses are not perturbative responses in general, unless constraints on the number of bands or the spectrum are imposed as for the low-energy models of chiral fermions.

To see which non-perturbative optical response is characterized by $\chi^{mn}$, we relate it to the first Chern number $c_1$ of $n$ and $m$ bands by
\begin{align}
\label{eq:Euler-Chern}
\chi^{mn}=c^n_1-c^m_1.
\end{align}
Since $c_1$ is proportional to the Hall conductivity, equation~\eqref{eq:Euler-Chern} shows that the photovoltaic Hall effect by the complete population inversion between $n$ and $m$ bands is an example of quantized optical responses due to $\chi^{mn}$.
Therefore, the complete population inversion by extreme optical pumping is a universal mechanism for topological optical responses.

The above features of topological optical responses appear in $PT$-symmetric systems also, where the reality or symplecticity condition is imposed on the geometry.
The relevant topological invariants are non-polynomial functions of geometric quantities and have relations similar to Eq.~\eqref{eq:Euler-Chern}.
See Methods 9 and Supplementary Notes 1-3 for details.
\\

{\bf \noindent Conclusion.}
Our interpretation of transition dipole moment matrix elements as tangent vectors provides a concrete conceptual ground for understanding quantum geometry of electromagnetic responses in a unified way.
This is because geometric responses to static electromagnetic fields originate from virtual transitions, which are also described by transition dipole moments (see Methods 10 for details).
For example, as we already note above, the Fubini-Study metric of a state $\ket{u_n}$ is given by summing the Hermitian metric $Q^{mn}$ over all possible intermediate state $\ket{u_m}$.
This unified geometric perspective helps us relate seemingly different physical phenomena.
Since the geometric superfluid weight and the linear optical conductivity are described by the quantum metric, one may expect that there exists a physical relation between the two.
Indeed, they are intimately related by the optical frequency sum rule~\cite{hazra2019bounds,verma2021optical,ahn2021superconductivity}.
Similar relations are expected for other geometric responses also.
Optical measurements are thus promising universal probes of quantum geometry, in practice with a reasonable cutoff on the number of intermediate states for static responses.

Moreover, our theoretical approach can be applied beyond electromagnetic responses to understand other geometric responses such as the photovoltaic thermal Hall and Seebeck effects.
It is also possible to study the real-space quantum geometry by identifying momentum operators as tangent vectors.
We leave these directions for future studies.
\\

\vspace{0.1in}
{\bf \large \noindent Acknowledgements}\\
{\small We appreciate Eslam Khalaf and Daniel Parker for helpful discussions and thank Maine Christos for useful comments on the manuscript.
J.A. was supported by the Basic Science Research Program through the National Research Foundation of Korea (NRF) funded by the Ministry of Education (Grant No. 2020R1A6A3A03037129).
J.A. and A.V. were supported by the Center for Advancement of Topological Semimetals, an Energy Frontier Research Center funded by the US Department of Energy Office of Science, Office of Basic Energy Sciences, through the Ames Laboratory under contract No. DE-AC02-07CH11358.
G.-Y. G. acknowledges the support from the Ministry of Science and Technology and National Center for Theoretical Sciences in Taiwan and thanks the National Center for High-performance Computing in Taiwan for the computing time.
N.N. was supported by JST CREST Grant Number JPMJCR1874 and JPMJCR16F1, Japan, and JSPS KAKENHI Grant Number 18H03676.
}
\\

{\bf \large \noindent Author contributions}\\
{\small 
J.A. conceived the original idea and performed theoretical analysis.
G.-Y.G. performed first-principles calculations.
N.N. and A.V. supervised the project.
All authors discussed results and contributed to the formulation of theory and writing of the manuscript.
}
\\

{\bf \large \noindent Competing interests}\\
{\small The authors declare no competing financial interests.}
\\

{\bf \large \noindent Additional information}\\
{\small {\bf \noindent Correspondence and requests for materials} may be addressed to any of the authors.}
\\

\clearpage
\newpage


\bibliographystyle{naturemag}

{\bf \noindent \large Methods}\\
{\bf \noindent 1. Conventions and Definitions.}
\begin{itemize}

\item $-e$ is the electron charge. $e>0$.

\item
Typical quantities in optical conductivity tensors.
\begin{align}
\d_a
&=\frac{\d}{\d_{k_a}},\notag\\
\int_{\bf k}
&=\int \frac{d^dk}{(2\pi)^d},\notag\\
\hat{H}_0
&=\text{single-electron Hamiltonian without external field}\notag\\
E_n
&=\braket{\psi_{n\bf k}|\hat{H}_0|\psi_{n\bf k}}=\braket{u_{n\bf k}|\hat{H}_0({\bf k})|u_{n\bf k}},\notag\\
v^a_{mn}
&=\frac{i}{\hbar}\braket{\psi_{m\bf k}|[\hat{r}^a,\hat{H}_0]|\psi_{n\bf k}}
=\hbar^{-1}\braket{u_{m\bf k}|\d_a\hat{H}_0({\bf k})|u_{n\bf k}},\notag\\
f_{m}
&=\text{Fermi-Dirac distribution of the $m$th band},\notag\\
\omega_{mn}
&=\hbar^{-1}\left(E_m-E_n\right),\notag\\
f_{mn}
&=f_m-f_n,\notag\\
r^a_{mn}
&=i\delta_{mn}\d_a+\braket{u_{m\bf k}|i\d_a|u_{n\bf k}},\notag\\
({\cal A}_a)_{mn}
&=\braket{u_{m\bf k}|i\d_a|u_{n\bf k}}\delta_{E_m,E_n}\notag\\
{\cal D}_a
&=\d_a-i{\cal A}_a,\notag\\
r^a_{mn;c}
&=[{\cal D}_c,r^a]_{mn},\notag\\
{\cal F}_{ba}
&=i[{\cal D}_b,{\cal D}_a]
=\d_b{\cal A}_a-\d_a{\cal A}_b-i[{\cal A}_b,{\cal A}_a].
\end{align}

\item
Definition of geometric quantities in momentum space defined for a pair of energy levels $E_n$ and $E_m$.
We consider degenerate states $\ket{u_{n_i}}$s with $E_{n_i}=E_n$ and $\ket{u_{m_j}}$s with $E_{m_j}=E_{m}$.

\begin{align}
Q_{ba}
&=\sum_{i,j}r^b_{n_im_j}r^a_{m_jn_i}\equiv g_{ba}-iF_{ba}/2,\notag\\
C_{abc}
&=\sum_{i,j}r^b_{n_im_j}r^a_{m_jn_i,c}\equiv \Gamma_{bca}-i\tilde{\Gamma}_{bca},\notag\\
K_{badc}
&=\sum_{i,j}r^b_{n_im_j}r^a_{m_jn_i,dc}-r^b_{nm}r^a_{mn,cd}\equiv R_{badc}-i\tilde{R}_{badc},\notag\\
D_{badc}
&=\sum_{i,j}r^b_{n_im_j}r^a_{m_jn_i,dc}.
\end{align}

\item
Name of geometric quantities
\begin{align}
&Q_{ba}=g_{ba}-iF_{ba}/2						&\text{Hermitian metric}\notag\\
&g_{ba}										&\text{Riemannian metric}\notag\\
&F_{ba}/2									&\text{symplectic form}\notag\\
&C_{bca}=\Gamma_{bca}-i\tilde{\Gamma}_{bca}&\text{Hermitian connection}\notag\\
&\Gamma_{bca} 								&\text{metric connection}\notag\\
&\tilde{\Gamma}_{bca} 						&\text{symplectic connection}\notag\\
&K_{badc}=R_{badc}-i\tilde{R}_{badc}			&\text{Hermitian curvature}\notag\\
&R_{badc} 								&\text{Riemann curvature tensor}\notag\\
&\tilde{R}_{badc} 							&\text{symplectic curvature tensor}\notag\\
&D_{badc}								&\text{second-order connection}
\end{align}

\end{itemize}

{\bf \noindent 2. Nonlinear conductivity tensors.}
By perturbatively expanding the current expectation value in terms of external electric fields, one obtains
\begin{align}
j^{a}(t)
&=-e
\sum_k
\sum_{a_1,\hdots,a_k}
\left[\prod_{i=1}^{k}\int^{t_{i+1}}_{-\infty}\frac{ie}{\hbar}E^{a_i}(t_i)\right]\notag\\
&\qquad\times\braket{[\hat{r}^{a_{k}},\hdots[\hat{r}^{a_1},\hat{v}^{a}]]},
\end{align}
where $t_{k+1}=t$ at each order $k$.
The Fourier components of the $k$th-order optical conductivity tensors $\tilde{\sigma}^{a;a_1,\hdots,a_k}$ are defined by 
\begin{align}
j^{a}(\Omega)
&=
\sum_k
\sum_{\substack{a_1,\hdots,a_k\\ \omega_1,\hdots,\omega_k}}E^{a_1}(\omega_1)\hdots E^{a_k}(\omega_k)\notag\\
&\qquad\times\tilde{\sigma}^{a;a_1,\hdots,a_k}(\Omega;\omega_1,\hdots,\omega_k)\delta_{\Omega,\sum_{i=1}^k\omega_i},
\end{align}
where we Fourier transform the current and electric fields using $f(t)=\sum_{\omega}f(\omega)e^{-i\omega t}$, and the Fourier component satisfies $f(-\omega)=[f(\omega)]^*$ because $f(t)$ we consider is real valued.
The conductivity tensor (unsymmetrized) has the form~\cite{ventura2017gauge}
\begin{align}
&\tilde{\sigma}^{a;a_1,\hdots,a_k}\left(\Omega,\omega_1,\hdots,\omega_k\right)=-e\left(\frac{ie}{\hbar}\right)^k\times\notag\\
&\int_{\bf k}
\sum_n
f_n\bra{\psi_{n\bf k}}
[\hat{r}^{a_k},G_k
\circ
\hdots 
[\hat{r}^{a_{1}},G_{1}
\circ v^{a_{k+1}}]]
\ket{\psi_{n\bf k}},
\end{align}
where $f_n$ is the Fermi-Dirac distribution of the $n$th band, $\braket{\psi_{m\bf k'}|\hat{\bf r}^a|\psi_{n\bf k}}=\delta_{mn}i\d_a\delta_{{\bf k}',{\bf k}}+\braket{u_{m\bf k'}|i\d_a|u_{n\bf k}}$ in momentum space,
$A\circ B = \sum_{\alpha,\beta}A_{\alpha \beta}B_{\alpha \beta}$ is the Hadamard product in the Hilbert space, and $G_i=[-i(\hat{\omega}+\sum_{i=1}^{k}\omega_i)]^{-1}$, where $\braket{\psi_{m\bf k'}|\hat{\omega}|\psi_{n\bf k}}=\omega_{mn}\delta_{\bf k',k}$, and $\omega_{mn}=\hbar^{-1}(E_m-E_n)$.
When the first operator $\hat{r}^{a_k}$ acts as a derivative, it produces total derivative terms, which are Fermi surface contributions.
The other terms are inter-band contributions.
 
In this work, we use the symmetrized conductivity tensor defined by~\cite{boyd2020nonlinear}
\begin{align}
&\sigma^{a;a_1,\hdots,a_k}\left(\Omega,\omega_1,\hdots,\omega_k\right)\notag\\
&=\frac{1}{N_P}\sum_{\rm Perm}\tilde{\sigma}^{a;a_1,\hdots,a_k}\left(\Omega,\omega_1,\hdots,\omega_k\right),
\end{align}
where $\sum_{\rm Perm}$ is the summation over all permutations of the $k$ pairs $(a_i,\omega_i)$ for $i=1,\hdots,k$, and $N_P=\sum_{\rm Perm}1$ is the number of all possible permutations.
\\
 
{\bf \noindent 3. Photoconductivity tensors.}
We are interested in the third-order photoconductivity described by
\begin{align}
j^d(0)
&=\left[\sigma^{d;a}_{\rm dark}+\sigma^{d;a}_{\rm photo}(\omega)\right]E^a(0),
\end{align}
where $\sigma^{d;a}_{\rm dark}$ is the linear conductivity without incident light,
\begin{align}
\sigma^{d;a}_{\rm photo}(\omega)
=
6\sum_{b,c}\sigma^{d;abc}(0,0,\omega,-\omega)
E^b(\omega)E^c(-\omega)
\end{align}
is the photoconductivity, and
\begin{align}
&\sigma^{d;abc}(0;0,\omega,-\omega)\notag\\
&=\frac{1}{6}\bigg[
\tilde{\sigma}^{d;abc}(0;0,\omega,-\omega)
+\tilde{\sigma}^{d;acb}(0,0,-\omega,\omega)\notag\\
&+\tilde{\sigma}^{d;cab}(0;-\omega,0,\omega)
+\tilde{\sigma}^{d;bac}(0;\omega,0,-\omega)\notag\\
&+\tilde{\sigma}^{d;bca}(0;\omega,-\omega,0)
+\tilde{\sigma}^{d;cba}(0;-\omega,\omega,0)\bigg]
\end{align}
is the symmetrized third-order photoconductivity tensor.
It consists of jerk, injection, and shift responses that are proportional to $\Gamma^{-2}$, and $\Gamma^{-1}$, and $\Gamma^{0}$~\cite{fregoso2019bulk}.
In the clean limit where the relaxation energy scale $\hbar\Gamma$ is much smaller than the photon energy $\hbar\omega$ and the band gaps $\hbar\omega_{mn}$s, the jerk response is the most dominant response.
However, it does not generate Hall conductivity because $\sigma^{d;abc}_{\rm jerk}\left(0;0,\omega,-\omega\right)
=\frac{\pi e^4}{\hbar^3}
\frac{1}{6\Gamma^2}
\int_{\bf k}
\sum_{n,m}
f_{nm}
\delta(-\omega+\omega_{mn})
2r^{c}_{nm}r^b_{mn}\d_a\d_d\omega_{mn}$
is symmetric with respect to indices $a$ and $d$.
Accordingly, the most dominant photovoltaic Hall conductivity comes from the injection response.
\begin{align}
\label{eq:PVH-general}
&\sigma^{d;abc}_{\rm inj}\left(0;0,\omega,-\omega\right)
=\frac{\pi e^4}{\hbar^3}
\frac{1}{6\Gamma}\int_{\bf k}
\sum_{n,m}
f_{nm}\delta(-\omega+\omega_{mn})\times\notag\\
&
i\Bigg\{
K^{mn}_{cbad}
+D^{mn}_{cb(ad)}
-(D^{mn}_{bc(ad)})^*\notag\\
&+\sum_e\left((C^{mn}_{edc})^*(C^{mn})^e_{ab}
-(C^{mn}_{eac})^*(C^{mn})^e_{db}\right)
\notag\\
&+2\frac{\d_d\omega_{mn}}{\omega_{mn}}
\left[
C^{mn}_{cba}
-(C^{mn}_{bca})^*
\right]\notag\\
&-2\frac{\d_d\omega_{mn}}{\omega_{mn}}\left(
Q^{mn}_{ca}\frac{\d_b\omega_{mn}}{\omega_{mn}}
-(Q^{mn}_{ba})^*\frac{\d_c\omega_{mn}}{\omega_{mn}}\right)
\notag\\
&+2\d_d\omega_{mn}
\sum_{p:E_p\ne E_m,E_n}
\bigg[r^{c}_{nm}
\left(
r^b_{mp}\frac{r^a_{pn}}{\omega_{pn}}
-\frac{r^a_{mp}}{\omega_{mp}}r^b_{pn}
\right)\notag\\
&+\left(
r^{c}_{np}\frac{r^a_{pm}}{\omega_{pm}}
-\frac{r^a_{np}}{\omega_{np}}r^{c}_{pm}
\right)
r^b_{mn}
\bigg]
\Bigg\},
\end{align}
where $K^{mn}_{cbad}=r^c_{nm}(r^b_{mn,da}-r^b_{mn,ad})$, $D^{mn}_{cb(ad)}=\frac{1}{2}r^c_{nm}(r^b_{mn,da}+r^b_{mn,ad})$, $Q^{mn}_{ba}=r^b_{nm}r^a_{mn}$, $\sum_eQ^{mn}_{ce}(C^{mn})^e_{ba}=C^{mn}_{cba}=r^c_{nm}r^a_{mn,b}$, $\sum_e(C^{mn}_{edc})^*(C^{mn})^e_{ab}=r^c_{nm,d}r^b_{mn,a}$, $O_{mn,a}=\d_aO_{mn}-i[{\cal A},O]_{mn}$, and ${\cal A}_{mn}=\delta_{E_m,E_n}\braket{u_{m}|\d_a|u_{n}}$ is the Berry connection.
$\delta_{E_m,E_n}=\delta_{mn}$ when bands are all non-degenerate, as we assume in the most of part of the main text.
\\

{\bf \noindent 4. Maurer-Cartan form.}
Here we explain the basic concept of the Maurer-Cartan form (See Ref.~\cite{nakahara2003geometry} for further information).
Let us consider the space of complex $N\times N$ matrices ${\rm M}(N,{\bb C})$, defining an $N^2$-dimensional complex Euclidean space ${\bb C}^{N^2}$.
The geometrical structure of subspaces such as ${\rm U}(N)$ or ${\rm U}(N)/{\rm U}(1)^N$ is induced from the structure of ${\rm M}(N,{\bb C})$.
Given a matrix representation, we can define Euclidean coordinates $x^{ij}$ with $i,j=1,\hdots,N$ by
\begin{align}
\label{eq:matrix-coordinates}
x^{ij}(g)=g^{ij},\quad g\in {\rm M}(N,{\bb C}),
\end{align}
where $g^{ij}$ is the matrix element of $g$.
The tangent space at $g$ is spanned by $\frac{\d}{\d x^{ij}}|_{g}$.
Here the symbol $|_{g}$ indicate that it is defined at the point $g\in {\rm M}(N,{\bb C})$.
Two tangent basis vectors $\frac{\d}{\d x^{ij}}|_{g}$ and $\frac{\d}{\d x^{ij}}|_{g'}$ are different objects when $g\ne g'$ because they live at distinct points.

Let us restrict our attention to the space of invertible matrices ${\rm GL}(N,{\bb C})$.
The left multiplication $L_f:g\rightarrow fg$ for invertible matrices $f$ and $g$ induces a push-forward mapping $L_{f*}$ between their tangent vectors
\begin{align}
L_{f*}\d_{x^{ij}}|_{g}
=\frac{\d x^{kl}(fg)}{\d x^{ij}(g)}\d_{x^{kl}}|_{fg}
\end{align}
Using the property of the coordinates we choose in equation~\eqref{eq:matrix-coordinates}, we have
\begin{align}
L_{f*}X|_g
&=\sum_{k,j}\left[f X(g)\right]^{kj}\d_{x^{kj}(fg)}
\end{align}
for any vector field $X$, where $X|_g=\sum_{i,j}X^{ij}(g)\d_{x^{ij}(g)}$ is a tangent vector at $g$ defined by $X$, and $(fX)^{kl}=\sum_mf^{km}X^{ml}$ is the matrix product.
If we take $f=g^{-1}$ and consider the vector field describing the tangent basis vectors obtained by momentum space coordinates $k^a$s, i.e., $X(g)=\d_a|_{g}=\sum_{ij}\left[\d_ax^{ij}(g)\right]\d_{x^{ij}}(g)$,
\begin{align}
L_{g^{-1}*}\left[\sum_{i,j}(\d_ag^{ij})\d_{x^{ij}(g)}\right]
&=\sum_{k,j}\left(g^{-1} \d_ag\right)^{kj}\d_{x^{kj}(1)},
\end{align}
where $1$ is the identity matrix.
This shows that the Maurer-Cartan form $\xi_a=g^{-1} \d_ag$ is the tangent vector at $g$ that is sent back to the Lie algebra ${\rm gl}(N,{\bb C})$, which is the tangent space defined at the identity matrix.
More formally, as a differential form, the Maurer-Cartan form is defined as a linear map from tangent vectors  to the Lie algebra by $\xi (X|_g)\equiv L_{g^{-1}*}X|_g$, from which $\xi (\d_a|_g)=g^{-1}\d_ag$.
We simply call the Lie algebras $g^{-1}\d_ag$s Maurer-Cartan forms as is typical.

The Lie bracket of the Maurer-Cartan forms is evaluated using the matrix commutator. $[V,W]^{ij}_{\rm LB}=-(V^{ik}W^{kj}-W^{ik}V^{kj})=-[V,W]^{ij}$, where $V=g^{-1}\d_bg$ and $W=g^{-1}\d_ag$.
\\

{\bf \noindent 5. Riemannian geometry with degenerate states.}
Let us consider two distinct energy levels $E_n$ and $E_m$ with $N_d$- and $M_d$-fold degenerate states $\ket{u_{n_i}}$ and $\ket{u_{m_j}}$, respectively, where $i=1,\hdots, N_d$ and $j=1,\hdots,M_d$.
Then, the tangent vector associated with the momentum coordinate $k^a$ is defined by
\begin{align}
\hat{e}^{mn}_a
\equiv \sum_{i,j}r^a_{m_jn_i}\ket{u_{m_j}}\bra{u_{n_i}}.
\end{align}
The Hermitian metric is evaluated from the Hilbert-Schmidt inner product of two such basis vectors.
\begin{align}
Q^{mn}_{ba}
&\equiv (\hat{e}^{mn}_b,\hat{e}^{mn}_a)\notag\\
&=\sum_{ij}r^b_{n_im_j}r^a_{m_jn_i}.
\end{align}

The covariant derivative is obtained by projecting to the basis $\ket{u_{m_j}}\bra{u_{n_i}}$s after taking an ordinary derivative.
\begin{align}
\nabla_c\hat{e}^{mn}_a
\equiv P_m(\d_c\hat{e}_a)P_n
= \sum_{i,j}r^a_{m_jn_i,c}\ket{u_{m_j}}\bra{n_i},
\end{align}
where $P_m=\sum_{j}\ket{u_{m_j}}\bra{u_{m_j}}$, $P_n=\sum_{i}\ket{u_{n_i}}\bra{u_{n_i}}$, and
\begin{align}
r^a_{m_jn_i,c}
=[{\cal D}_c,r^a]_{m_jn_i},
\end{align}
where $[A,B]=AB=BA$ is the commutator, ${\cal D}_c=\d_c-i{\cal A}^c$, and ${\cal A}_{pq}=\delta_{E_{p},E_{q}}\braket{u_p|\d_c|u_q}$ is the ${\rm U}(N_d)\times {\rm U}(M_d)$ Berry connection [${\rm U}(N_d)$ for $n$ and ${\rm U}(M_d)$ for $m$, and forget about the gauge group of the other states here].
The Hermitian connection is then defined by
\begin{align}
C^{mn}_{bca}
\equiv (\hat{e}^{mn}_b,\nabla_c\hat{e}^{mn}_a)
=\sum_{ij}r^b_{n_im_j}r^a_{m_jn_i,c},
\end{align}
This connection satisfies the metric compatibility $\nabla_c(\hat{e}^{mn}_b,\hat{e}^{mn}_a)=(\hat{e}^{mn}_b,\nabla_c\hat{e}^{mn}_a)+(\nabla_c\hat{e}^{mn}_b,\hat{e}^{mn}_a)$ because $(\hat{e}^{mn}_b,\nabla_c\hat{e}^{mn}_a)=(\hat{e}^{mn}_b,\d_c\hat{e}^{mn}_a)$, and $\nabla_c$ acts as an ordinary derivative on scalars.
Here, the generalized derivative preserves the Hermiticity $(O_{mn,c})^*=O_{nm,c}$ for a Hermitian operator $O$.

The Hermitian connection we define has a non-zero torsion in general when the number of bands exceeds two.
The torsion tensor is given by virtual transitions among three bands.
\begin{align}
T^{mn}_{bca}
&\equiv (\hat{e}^{mn}_b,\nabla_c\hat{e}^{mn}_a-\nabla_a\hat{e}^{mn}_c-i[\hat{e}^{mn}_c,\hat{e}^{mn}_a])\notag\\
&=C^{mn}_{bca}-C^{mn}_{bac}\notag\\
&=i\sum_{i,j}r^b_{n_im_j}\sum_{p:E_p\ne E_m,E_n}\left(r^c_{m_jp}r^a_{pn_i}-r^a_{m_jp}r^c_{pn_i}\right),
\end{align}
where we use that $[\hat{e}^{mn}_c,\hat{e}^{mn}_a]$=0.

The Hermitian curvature tensor is defined by
\begin{align}
K^{mn}_{badc}
\equiv (\hat{e}^{mn}_b,(\nabla_d\nabla_c-\nabla_c\nabla_d-\nabla_{[\hat{e}^{mn}_d,\hat{e}^{mn}_c]})\hat{e}^{mn}_a).
\end{align}
It is identical to $D^{mn}_{badc}-D^{mn}_{badc}$ because $[\hat{e}^{mn}_d,\hat{e}^{mn}_c]=0$, where
\begin{align}
D^{mn}_{badc}
&\equiv (\hat{e}^{mn}_b,\nabla_d\nabla_c\hat{e}^{mn}_a)
=\sum_{ij}^{N_d}r^b_{n_im_j}r^a_{m_jn_i,cd}
\end{align}
is the second-order connection, and
\begin{align}
r^a_{m_jn_i,cd}
&=[{\cal D}_d,[{\cal D}_c,r^a]]_{m_jn_i}
\end{align}
is the double covariant derivative.
The index $c$ of the double covariant derivative does not transform tensorially under general coordinate transformations, in contrast to case of the second covariant derivative $S^{mn}_{badc}=(\hat{e}^{mn}_b,(\nabla_d\nabla_c-\nabla_{\nabla_{d}\hat{e}^{mn}_c})\hat{e}^{mn}_a)$.
However, the double covariant derivative is more convenient for our purpose.
The Hermitian curvature tensor can be written as
\begin{align}
\label{eq:H-curvature-Berry}
K^{mn}_{badc}
&=-i\sum_{i,j}r^b_{n_im_j}[{\cal F}_{dc},r^a]_{m_jn_i},
\end{align}
where ${\cal F}_{dc}$ is the Berry curvature $({\cal F}_{dc})_{pq}=i[{\cal D}_d,{\cal D}_c]_{pq}=\left(\d_d{\cal A}_c-\d_c{\cal A}_d-i[{\cal A}_d,{\cal A}_c]\right)_{pq}.$
It follows from
\begin{align}
r^a_{m_jn_i,cd}-r^a_{m_jn_i,dc}
&=-i[{\cal F}_{dc},r^a]_{m_jn_i},
\end{align}
where we use $[{\cal D}_d,[{\cal D}_c,r^a]]_{m_in_j}-[{\cal D}_c,[{\cal D}_d,r^a]]_{m_jn_i}=[[{\cal D}_d,{\cal D}_c],r^a]_{m_jn_i}$.
A useful property of the Hermitian curvature is
\begin{align}
(K^{mn}_{badc})^*
=K^{mn}_{abdc},
\end{align}
which can be shown from equation~\eqref{eq:H-curvature-Berry} and the Hermitian property $(K^{mn}_{badc})^*=K^{nm}_{badc}$.
\\

{\bf \noindent 6. Symmetry constraints on the geometric quantities.}
To study symmetry properties, it is convenient to write $\hat{e}^{mn}_a$ using projection operators as
$\hat{e}^{mn}_a({\bf k})
=P_m({\bf k})i\d_{k_a}P_n({\bf k})$,
where $P_n({\bf k})=\sum_{i=1}^{N_d}\ket{u_{n_i\bf k}}\bra{u_{n_i\bf k}}$ is the projection to $N_d$-fold degenerate states with $E_{n_i}=E_n$.

Let us first consider time reversal symmetry.
It imposes $TP_m({\bf k})T^{-1}=P_m(-{\bf k})$, where $T$ is the time reversal operator acting on the cell-periodic Bloch states.
Accordingly, $e_a$ follows the corresponding symmetry constraint.
$T\hat{e}^{mn}_a({\bf k})T^{-1}=P_m({-\bf k})i\d_{-k_a}[P_n(-{\bf k})]=\hat{e}^{mn}_a(-{\bf k})$.
Using this, one can show that time reversal symmetry imposes
\begin{align}
G({\bf k})=G^*(-{\bf k})
\end{align}
for $G=Q^{mn}$, $C^{mn}$, $D^{mn}$, or $K^{mn}$

Spatial symmetry under $g:{\bf r}\rightarrow R_g{\bf r}+{\bf a}_g$ imposes $U_gP_m({\bf k})U_g^{-1}=P_m(R_g{\bf k})$, where $U_g$ is the unitary representation of $g$ for cell-periodic Bloch states.
The $g$ symmetry constraint on the tangent vector is $U_g\hat{e}^{mn}_a({\bf k})U_g^{-1}=P_m({R_g\bf k})i\d_{k_a}[P_n(R_g{\bf k})]=\sum_b(R_g)_{ba}\hat{e}^{mn}_b(-{\bf k})$.
Geometric quantities then satisfy
\begin{align}
G_{a_1\hdots a_k}({\bf k})=(R_g)_{b_1a_1}\hdots (R_g)_{b_ka_k}G_{b_1\hdots b_k}(R_g{\bf k})
\end{align}
for $G=Q^{mn}$, $C^{mn}$, $D^{mn}$, or $K^{mn}$.

Spacetime inversion $PT$ symmetry plays a distinguished role because it gives constraints at each ${\bf k}$ point.
The constraint $PT\hat{e}^{mn}_a({\bf k})(PT)^{-1}=\hat{e}^{mn}_a({\bf k})$ gives a real structure when $(PT)^2=1$.
This is most clearly seen by taking a gauge where $PT$ is represented by a complex conjugation $K$, such that $(\hat{e}^{mn}_a)^*=\hat{e}^{mn}_a$ is real-valued.
It forms a one-dimensional real subspace of the tangent space of the manifold ${\rm O}(N)/{\rm O}(1)^N$ of non-degenerate $PT$-symmetric cell-periodic Bloch states.
On the other hand, $PT$ symmetry gives a quaternion structure when $(PT)^2=-1$.
To see this, let us write down the matrix element of $\hat{e}^{mn}_a$.
Each band becomes twofold degenerate by the $PT$ symmetry satisfying $(PT)^2=-1$ (Kramers theorem).
We can fix the gauge such that $PT$ is represented by $i\sigma_yK$ within each set of twofold degenerate states.
Then $PT$ symmetry $(r^a)_{m_in_j}^*=-(i\sigma_y^{-1}r^ai\sigma_y)_{m_in_j}$ requires that $\left(r^a_{m_in_j}\right)=-i(f^a_1\sigma_0+f^a_2i\sigma_x+f^a_3i\sigma_y+f^a_4i\sigma_z)$,
where $\sigma_0$ is the $2\times 2$ identity matrix, and $\sigma_{i=x,y,z}$ are Pauli matrices, and $f^a_{i=1,2,3,4}$ are real-valued functions.
The matrices $\{\sigma_0,i\sigma_x,i\sigma_y,i\sigma_z\}$ satisfy the quaternion algebra, defining a quaternion structure.
They form a one-dimensional quaternionic subspace of the tangent space of the manifold ${\rm Sp}(N)/{\rm Sp}(1)^N$ of Kramers degenerate $PT$-symmetric cell-periodic Bloch states.
Independent of the sign of $(PT)^2$, the geometric quantities satisfy $G({\bf k})=G^*({\bf k})$ for $G=Q^{mn}$, $C^{mn}$, $D^{mn}$, or $K^{mn}$.
\\

{\bf \noindent 7. Riemann curvature tensor of two-level systems.}
Let us consider a model with two energy levels $E_n$ and $E_m$ which are $N_d$- and $M_d$-fold degenerate, respectively.
Using that
\begin{align}
({\cal Q}^n_{ba})_{n_in_k}=\sum_{j=1}^{M_d}r^b_{n_im_j}r^a_{m_jn_k}
\end{align}
defines the non-abelian Hermitian metric of the level $n$ (similarly for the level $m$), where $E_{n_{i}}=E_{n_k}=E_n$ and $E_{m_j}=E_m$, we obatin
\begin{align}
\label{eq:K_two-level}
K^{mn}_{badc}
&=i{\rm Tr}_n{\cal Q}^n_{ba}{\cal F}^n_{dc}-i{\rm Tr}_m{\cal F}^m_{dc}{\cal Q}^m_{ab},
\end{align}
where ${\cal Q}^n_{ba}={\cal G}^n_{ba}-(i/2){\cal F}^n_{ba}$ is the non-abelian Hermitian metric of the level $E_n$, and ${\rm Tr}_{n}$ is the trace within the energy level $E_n$.
The non-abelian Riemannian metric ${\cal G}^n_{ba}$ and the Berry curvature ${\cal F}^n_{ba}$ are respectively symmetric and anti-symmetric with respect to the exchange of $a$ and $b$.

Now we suppose that the system has chiral $S$ symmetry $SH({\bf k})S^{-1}=-H({\bf k})$ or particle-hole-times-inversion $CP$ symmetry  $CPH({\bf k})(CP)^{-1}=-H^*({\bf k})$.
Those symmetries impose
\begin{align}
{\cal Q}^m_{ba}({\bf k})
&=U_S^{-1}{\cal G}^n_{ba}({\bf k})U_S,\notag\\
{\cal Q}^m_{ba}({\bf k})
&=U_{CP}^{-1}[{\cal Q}^n_{ba}({\bf k})]^*U_{CP}.
\end{align}
The expression of the Hermitian curvature tensor then reduces to,
\begin{align}
K^{mn}_{badc}
&=2i{\rm Tr}_n{\cal Q}^n_{ba}{\cal F}^n_{dc}.
\end{align}
This formula can be applied to the linearized model of Dirac or Weyl point whether it is massive or massless, because such a model has $CP$ or chiral symmetry.
One interesting special case is ${\rm Re}K^{mn}_{baba}=R^{mn}_{baba}={\rm Tr}_n{\cal F}^n_{ba}{\cal F}^n_{ba}$, which measures the norm of the non-abelian Berry curvature matrix ${\cal F}^n_{ba}$.
\\

{\bf \noindent 8. Third-order photovoltaic Hall conductivity of Dirac and Weyl fermions.}
Let us consider a two-dimensional Dirac fermion described by equation~\eqref{eq:2DDirac}.
The Riemann curvature term is
\begin{align}
{\rm Im}\sigma^{[x;y]xy}_{K}
&=\frac{e^4}{12\hbar^3}\frac{1}{\Gamma}\frac{1}{\omega}\left(\frac{v}{\omega}\right)^2
\left(\frac{2m}{\hbar\omega}\right)^2
\Theta(|\omega|-2|m|),
\end{align}
while the remaining term has the opposite sign
\begin{align}
&{\rm Im}\sigma^{[x;y]xy}-{\rm Im}\sigma^{[x;y]xy}_{K}\notag\\
&=-\frac{e^4}{12\hbar^3}\frac{1}{\Gamma}\frac{1}{\omega}\left(\frac{v}{\omega}\right)^2
\left[
1-\left(\frac{2m}{\hbar\omega}\right)^2
\right]\
\Theta(|\omega|-2|m|).
\end{align}
The Riemann curvature leads to the dominant photovoltaic Hall response near the band edge $\hbar|\omega|= 2|m|$.
Other contributions become dominant when $\hbar|\omega|>2\sqrt{2}|m|$, where the third-order conductivity changes sign.
When we take parameters used in the main text, $v=8\times 10^6\;{\rm m/s}$ and $2|m|=17\;{\rm meV}$, the peak value $1.2\times 10^{-15}{\rm Am^2V^{-3}}$ at the band edge.
It corresponds to the photoconductivity $\sigma^{x;y}_{\rm photo}= (e^2/h)\times 3.5\times 10^{-4} I_{\rm light}/({\rm 1\; Wcm^{-2}})$, where $e^2/h=3.874\times 10^{-5}\Omega^{-1}$ is the conductance quantum per spin.
This magnitude is the same as the photoconductivity for a massless Dirac fermion at $\hbar\omega=17\;{\rm meV}$, but the sign is opposite.

Nonzero components of the linear photovoltaic Hall conductivity tensor in two dimensions are
\begin{align}
\label{eq:LPVH-2DDirac}
&{\rm Re}\sigma^{[x;y]yy}
={\rm Re}\sigma^{[x;y]xx}\notag\\
&=\frac{e^4}{24\hbar^3}\frac{1}{\Gamma}\frac{1}{\omega}\left(\frac{v}{\omega}\right)^2\left(\frac{2m}{\hbar\omega}\right)
\left[1+\left(\frac{2m}{\hbar\omega}\right)^2\right]
\Theta(|\omega|-2|m|),
\end{align}
while $\sigma^{[x;y]xy}=\sigma^{[x;y]yx}=0$ by rotational symmetry.
Here, the whole response is from the symplectic curvature tensor, and no other contributions exist.

For a Weyl fermion in three dimensions described by
\begin{align}
H_{\rm Weyl}=\hbar v(k_x\sigma_x+k_y\sigma_y+k_z\sigma_z),
\end{align}
there is only one non-vanishing linearly independent component of the photovoltaic Hall conductivity tensor, which is
\begin{align}
{\rm Im}\sigma^{[x;y]xy}
&=-\frac{e^4}{72\pi\hbar^3}\frac{1}{\Gamma}\frac{1}{\omega}\left(\frac{v}{\omega}\right),
\end{align}
and ${\rm Im}\sigma^{[x;y]xy}_K=-{\rm Im}\sigma^{[x;y]xy}$.
Other non-zero components are related by $SO(3)$ rotational symmetries
\begin{align}
{\rm Im}\sigma^{[x;y]xy}={\rm Im}\sigma^{[yz];yz}={\rm Im}\sigma^{[zx];zx}
\end{align}
and the inherent relations ${\rm Im}\sigma^{[a;d]cb}=-{\rm Im}\sigma^{[d;a]cb}=-{\rm Im}\sigma^{[a;d]bc}$ by definition.
The linear photovoltaic Hall conductivity is zero is due to time reversal $T=i\sigma_yK$ symmetry of a Weyl fermion around the band-crossing point.

Massless Dirac fermions in three dimensions have the photovoltaic Hall conductivity tensors twice that of a Weyl fermion, as it consists of two copies of Weyl fermions.
When the mass gap $2m$ is introduced by
\begin{align}
H_{\rm Dirac}=\hbar v(k_x\tau_z\sigma_x+k_y\tau_z\sigma_y+k_z\tau_z\sigma_z)+m\tau_x,
\end{align}
the non-vanishing linearly independent photovoltaic Hall conductivity component is
\begin{align}
{\rm Im}\sigma^{[x;y]xy}
&=\frac{e^4}{36\pi\hbar^3}\frac{1}{\Gamma}\frac{1}{\omega}\left(\frac{v}{\omega}\right)\left[4\left(\frac{2m}{\hbar\omega}\right)^2-1\right]\Theta(|\omega|-2|m|),
\end{align}
and
\begin{align}
{\rm Im}\sigma^{[x;y]xy}_K
&=\frac{e^4}{36\pi\hbar^3}\frac{1}{\Gamma}\frac{1}{\omega}\left(\frac{v}{\omega}\right)\left[1+2\left(\frac{2m}{\hbar\omega}\right)^2\right]\Theta(|\omega|-2|m|).
\end{align}
Here, we use $R^{cv}_{yxyx}={\rm Tr}\left[({\cal F}^n_{xy})^2\right]=8(v/\omega_{cv})^4[(2\hbar vk_z)^2+(2m)^2]/(\hbar\omega_{cv})^2$ ($c$ and $v$ indicating the upper- and lower-energy states).
See Methods 8. Since the mass preserves $PT=i\tau_x\sigma_yK$, the linear photovoltaic Hall conductivity is zero.
It does not depend on the choice of the matrix representation.
Any Dirac mass term preserves $PT$ symmetry.
\\

{\bf \noindent 9. Topology of the transition matrix dipole moment.}
When the spatial dimension $d$ is even, the Euler number is given by the generalized Gauss-Bonnet theorem~\cite{allendoerfer1943gauss}
\begin{align}
\chi^{mn}
&=\oint {\wedge^d}dk\sqrt{g}\sum_{a_i,b_j}\frac{\epsilon^{a_1a_2\hdots a_{d-1}a_d}\epsilon^{b_1b_2\hdots b_{d-1}b_d}}{(2\pi)^{d/2}2^d(d/2)!g}\notag\\
&\qquad R^{mn}_{a_1a_2b_1b_2}\hdots R^{mn}_{a_{d-1}a_{d}b_{d-1}b_d},
\end{align}
where $\epsilon$ is the Levi-Civita symbol, $g\equiv \det g^{mn}$ is the determinant of the metric $g^{mn}$ with respect to its momentum indices that are implicit in this notation,
${\wedge^d}dk$ is the oriented integral measure, and
$R^{mn}_{a_1a_2b_1b_2}={\rm Re}K^{mn}_{a_1a_2b_1b_2}$ is the Riemann curvature tensor.

In two-dimensions, we have
\begin{align}
\chi^{mn}
&=\frac{1}{2\pi}\oint dk_1dk_2{\rm sgn}(F^{mn}_{12})\frac{R^{mn}_{1212}}{\sqrt{g}}\notag\\
&=\frac{1}{2\pi}\oint dk_1dk_2\frac{|F^{mn}_{12}|({\cal F}_{12}^n-{\cal F}_{12}^m)}{2\sqrt{g}}\notag\\
&=\frac{1}{2\pi}\oint dk_1dk_2({\cal F}_{12}^n-{\cal F}_{12}^m)\notag\\
&=c_1^n-c_1^m,
\end{align}
where we use $|F^{mn}_{12}|/2=g$ following from $\det Q^{mn}=Q^{mn}_{11}Q^{mn}_{22}-Q^{mn}_{12}Q^{mn}_{21}=0$.
Here, $F^{mn}_{12}\equiv -2{\rm Im}Q^{mn}_{12}$ should be distinguished from the Berry curvature ${\cal F}$ (except in two-band systems).
$\det Q^{mn}=0$ because we introduce two momentum coordinates while $Q^{mn}$ is defined on one-dimensional tangent space (cf. $\det g$ is nonzero because the target space is two-dimensional as a real space).
Here, the sign ${\rm sgn}(F^{mn}_{12})$ due to the oriented nature of the integral measure should not be discarded.
Without this factor, one obtains a non-quantized value as in Refs.~\cite{ma2013euler,tan2019experimental,zhu2019note,ma2020euler}

In spin-orbit coupled $PT$-symmetric systems, where $(PT)^2=-1$, the transition dipole moments $\hat{e}^{m_in_j}_a$s between two pairs of Kramers-degenerate states $\ket{u_{m_{i=1,2}}}$ and $\ket{u_{n_{j=1,2}}}$ with $E_{m_1}=E_{m_2}\ne E_{n_1}=E_{n_2}$ define four-dimensional tangent subspace.
Therefore, the Gauss-Bonnet theorem can be applied in a four-dimensional momentum space.
The result is
\begin{align}
\chi^{mn}
&=\oint dk_1dk_2dk_3dk_4\sum_{a_i,b_j}\frac{{\rm sgn}(\wedge d^4k)}{(2\pi)^22^42\sqrt{g}}
\epsilon^{a_1a_2a_3a_4}\epsilon^{b_1b_2b_3b_4}\notag\\
&\qquad R^{nm}_{a_1a_2b_1b_2}R^{nm}_{a_3a_4b_1b_4}\notag\\
&=\oint \frac{d^4k}{32\pi^2}\sum_{b_j}
\epsilon^{b_1b_2b_3b_4}\notag\\
&\qquad\sum_{i=1}^2\left[
({\cal F}_{b_1b_2}{\cal F}_{b_3b_4})_{n_in_i}
-({\cal F}_{b_1b_2}{\cal F}_{b_3b_4})_{m_im_i}
\right]
\notag\\
&=p_1^n-p_1^m,
\end{align}
where $p_1$ is the first Pontryagin number.
See Supplementary Note 2 for a derivation.

In spinless $PT$-symmetric systems, where $(PT)^2=1$, $\hat{e}^{mn}_a$ define a one-dimensional real tangent vector.
Since it is odd-dimensional, the generalized Gauss-Bonnet theorem does not apply.
Instead, we can define the first Stiefel-Whitney number associated with it by
\begin{align}
w_1^{mn}
&=\frac{1}{\pi}\oint d{\bf k} \cdot i\d_{\bf k} \log r^a_{mn} \mod 2.
\end{align}
It measures the orientability of the tangent vector $\hat{e}^{mn}_a$ over the one-dimensional momentum space.
$w_1^{mn}=0$ ($w_1^{mn}=1$) indicates that $\hat{e}^{mn}_a$ is orientable (non-orientable).
As we show in Supplementary Note 3,
\begin{align}
w_1^{mn}=w^n_1-w^m_1,
\end{align}
where $w^{n}_1$ and $w^{m}_1$ are the first Stiefel-Whitney numbers of bands $n$ and $m$, or equivalently, the Berry phases of bands $n$ and $m$ divided by $\pi$~\cite{ahn2018band,ahn2019stiefel}.
\\

{\bf \noindent 10. Geometry of a single state and others.}
A single state $\ket{u_{n}}$ lives on the complex projective manifold
\begin{align}
{\rm {\bb C}P}^{N-1}=\frac{{\rm U}(N)}{{\rm U}(1)\times {\rm U}(N-1)},
\end{align}
where $N$ is the number of all bands, ${\rm U}(N)$ describes the degrees of freedom of the cell-periodic Bloch state $U_{\alpha n}=\braket{{\bf r}_{\alpha}|u_{n}}$, ${\rm U}(1)$ is the phase rotation of the state $\ket{u_{n}}$, and ${\rm U}(N-1)$ is the unitary transformation of the other states.
The tangent vectors on this manifold is described by the Maurer-Cartan form
\begin{align}
\hat{\xi}^n_a
&=U^{-1}i\d_aU-U^{-1}i\d_aU|_{{\rm U}(1)}-U^{-1}i\d_aU|_{{\rm U}(N-1)}\notag\\
&\equiv \hat{e}^n_a+(\hat{e}^n)_a^{\dagger},
\end{align}
where $U^{-1}i\d_aU=\sum_{m,n}\ket{u_m}r^a_{mn}\bra{u_{n}}$ is the Maurer-Cartan form of ${\rm U}(N)$, $U^{-1}i\d_aU|_{{\rm U}(1)}=\ket{u_n}\braket{u_n|i\d_a|u_n}\bra{u_{n}}$ is its projection to the ${\rm U}(1)$ direction, $U^{-1}i\d_aU|_{{\rm U}(N-1)}=\sum_{m\ne n,p \ne n}\ket{u_m}\braket{u_m|i\d_a|u_p}\bra{u_{p}}$ is the projection to ${\rm U}(N-1)$ direction, and
\begin{align}
\hat{e}^n_a&=\sum_{m:m\ne n}\ket{u_m}r^a_{mn}\bra{u_{n}}.
\end{align}
The Hermitian metric tensor of the $n$th band is
\begin{align}
Q^{n}_{ba}
=(\hat{e}^n_b,\hat{e}^n_a)
=\sum_{m:m\ne n}r^b_{nm}r^a_{mn}.
\end{align}
This is the Fubini-Study metric.
Equivalently, it can be defined by $Q^n_{ba}=\frac{1}{2}\left[(\hat{\xi}_b,\hat{\xi}_a)-i(\hat{\xi}_b,i\hat{\xi}_a)\right]$.
In two-band systems, the Fubini-Study metric is identical to the metric in equation~\eqref{eq:optical-metric} because there is only one pair of bands $(m,n)$; this is the reason why the geometric understanding of optical responses in two-band systems was possible through the Fubini-Study metric in the previous works~\cite{de2017quantized,de2020difference,ahn2020low}.
The imaginary part of the Fubini-Study metric, the symplectic form, is identical to the Berry curvature of the band $n$.
The Hermitian connection, torsion, and curvature can be defined from the covariant derivative
\begin{align}
O_{mn,c}
&=\left[\d_c-i{\cal A}_c^{{\rm U}(1)}-i{\cal A}_c^{{\rm U}(N-1)},O\right]_{mn},
\end{align}
where $({\cal A}_c^{{\rm U}(1)})_{mp}=\delta_{mn}\delta_{pn}\braket{u_n|i\d_c|u_n}$ is the ${\rm U}(1)$ Berry connection of band $n$, $({\cal A}_c^{{\rm U}(N-1)})_{mp}=(1-\delta_{mn})(1-\delta_{pn})\braket{u_m|i\d_c|u_p}$ is the ${\rm U}(N-1)$ Berry connection of the other bands.

Note that the torsion tensor is zero because $r^a_{mn,c}-r^c_{mn,a}=\d_cr^a_{mn}-\d_ar^c_{mn}-i[r^c,r^a]_{mn}=0$.
It is due to the fact that we consider the tangent space of a K\"ahler manifold, which has zero torsion~\cite{nakahara2003geometry}.
Any of its complex submanifolds also has zero torsion because it is a K\"ahler manifold~\cite{nakahara2003geometry}.
Let us compare this with that $\hat{e}^{mn}_a$ defines a subspace of the tangent space of the projective space.
The non-zero torsion (see Methods 5) of our connection for $\hat{e}^{mn}_a$ is the manifestation that the subspace of a tangent space may not define a tangent space of a submanifold by a topological obstruction (see Supplementary Note 1), i.e., not every vector bundle of dimension $D$ defines a tangent bundle of dimension $D$.
The torsionless connection is called the Levi-Civita connection.

The manifold of the whole occupied states is described by the complex Grassmannian manifold
\begin{align}
{\rm Gr}_{\bb C}(N_{\rm occ},N)=\frac{U(N)}{{\rm U}(N_{\rm occ})\times U(N-N_{\rm occ})}.
\end{align}
In this case, $\hat{\xi}_a= \hat{e}_a+\hat{e}_a^{\dagger}$, where
\begin{align}
\hat{e}_a&=\sum_{m\in {\rm unocc}}\sum_{n\in {\rm occ}}\ket{u_m}r^a_{mn}\bra{u_{n}}.
\end{align}
This geometric structure appears in dc responses because, in the limit $\omega\ll \omega_{mn}$ for all $m$ and $n$, all excitation channels contribute to the response non-trivailly through the Lorentzian $i/(\omega-\omega_{mn}+i\Gamma)\sim i/\omega_{mn}\ne 0$, which on the other hand behaves as the delta function $\delta(\omega-\omega_{mn})$ that chooses a specific pair of bands $(m,n)$ for resonant frequencies.
The Hermitian metric and other quantities are defined similarly.
The torsion is zero.

Let us remark that the formula equation~\eqref{eq:K_two-level}, used for two-level systems, applies also to the Hermitian curvature of a single state or occupied states.
Instead of considering two energy levels $E_n$ and $E_m$, we just need to consider two bipartite states (a single state vs the rest, or the occupied states vs the unoccupied states) and the corresponding change in the gauge group.
For example, the Hermitian curvature tensor of the $n$th state is $K^n_{badc}=i{\cal Q}^n_{ba}{\cal F}^n_{dc}-i{\rm Tr}_{n'}{\cal Q}^{n'}_{dc}{\cal F}^{n'}_{ab}$, where $n'$ indicates the set of all states excluding the $n$th state.

We can also consider general complex flag manifolds
\begin{align}
{\rm Fl}_{\bb C}(N_1,\hdots,N_k)
=\frac{U(N)}{{\rm U}(1)\times\hdots\times {\rm U}(N_k)}
\end{align}
with $\sum_{i=1}^kN_i=N$.
These are all K\"ahler manifolds, so they have zero torsion tensor.
It is straightforward to calculate the geometric quantities of the flag manifolds and their real and quaternion counterparts.
One just needs to consider different gauge groups when the Maurer-Cartan form and the covariant derivative are defined.
Then, one can show that $T_{bca}=C_{bca}-C_{bac}-i(\hat{e}_b,[\hat{e}_c,\hat{e}_a])=0$ (see Supplementary Note 4).
\\

{\bf \noindent 11. First-principles calculations.}
We perform first-principles calculations based on the density functional theory with
the generalized gradient approximation (GGA)~\cite{perdew1996generalized}.
We consider monolayer germanene modelled by a slab-supercell, ferromagnetic (LaOsO$_3$)$_2$ bilayer in the (111) (LaOsO$_3$)$_2$/(LaAlO$_3$)$_{10}$ superlattice (see~\cite{chandra2017quantum} for the superlattice structure), and bulk Bi$_2$Se$_3$.
For germanene, the separation between two neighboring slabs 
is at least 15 \AA$ $ in order to minimize the artificial inter-slab interaction.
The electronic structure calculations are carried out using the accurate projector-augmented wave (PAW) method,
as implemented in the Vienna {\it ab-initio} simulation package (VASP)~\cite{kresse1993ab,kresse1996efficient}.
The fully relativistic PAW potentials are adopted in order to include the spin-orbit coupling (SOC) effect.
Large plane-wave cutoff energies of $450 \;\rm eV$, $400\;\rm eV$, and 350 $\rm eV$ are used for germanene, (LaOsO$_3$)$_2$ bilayer, and Bi$_2$Se$_3$, respectively.
For the Brillouin zone (BZ) integration, a $k$-point mesh of $20 \times 20\times 1$, $12 \times 12 \times 2$ and $12 \times 12\times 12$ is used respectively.
All the self-consistent electronic structure calculations are performed with an energy convergence 
within $10^{-6}{\;\rm eV}$ between the successive iterations.

The third-order injection photovoltaic Hall conductivity is calculated 
using Eq.~\eqref{eq:PVH-general} in Methods 3. Since a large number of $k$-points are needed to get accurate NLO responses~\cite{ahn2020low},
we use the efficient Wannier function interpolation method based on maximally localized Wannier functions
(MLWFs)~\cite{wang2006ab,marzari2012maximally,ibanez2018ab}.
MLWFs of $p$ orbitals for the Ge, Bi and Se atoms and $d$ orbitals for the Os atom are constructed 
by fitting to the GGA+SOC band structures [Fig.~3(a,b)]. 
The band structures generated by the Wannier function interpolation are indistinguishable from
the corresponding GGA+SOC band structures. 
The third-order injection current conductivity spectra [Fig.~3(c,d)] are evaluated by
using dense $k$-meshes of $10000\times 10000\times 1$, $300\times 300\times 50$, and $500\times 500\times 500$ for germanene, (LaOsO$_3$)$_2$ bilayer, and Bi$_2$Se$_3$, respectively.
We find that the third-order conductivity obtained using such dense $k$-point meshes converge within a few percent.
Here we consider the "cold" materials, i.e., the Fermi-Dirac function in Eq.~\eqref{eq:PVH-general} is taken to be a step function.
Furthermore, the Dirac $\delta$ function is replaced by a Gaussian function with broadening width
of $\varepsilon = 5\; {\rm meV}$.
We use the relaxation rate $\Gamma$ given by $\hbar\Gamma = 1{\rm \; meV}$.\\

{\bf \large \noindent Data Availability}\\
{\small The data that support the findings of this study are available from the corresponding author upon reasonable request.}\\

{\bf \large \noindent Code Availability}\\
{\small The code that support the findings of this study are available from the corresponding author upon reasonable request.}

\clearpage
\newpage

\renewcommand{\thefigure}{S\arabic{figure}}
\renewcommand{\theequation}{S\arabic{equation}}
\renewcommand{\thetable}{S\arabic{table}}
\renewcommand{\thesubsection}{Supplementary Note \arabic{subsection}}

\setcounter{figure}{0} 
\setcounter{equation}{0} 
\setcounter{table}{0} 
\setcounter{section}{0}

\subsection{Obstruction to finding an associated submanifold}

Here, we show that there can be a topological obstruction for $\hat{e}^{mn}_a$ to define a submanifold of ${\cal M}={\rm U}(N)/{\rm U}(1)^N$.
Let us first note that the complex projective line ${\rm {\bb C}P}^1\simeq S^2$ is the only compact and closed one-dimensional complex manifold.
Since ${\rm {\bb C}P}^1$ has the Euler characteristic of two, the Euler characteristic calculated in the Brillouin zone should be $\chi^{mn}=N_w\chi({\rm {\bb C}P}^1)=2N_w$, where $N_w$ is the winding number of the map from the Brillouin zone to ${\rm {\bb C}P}^1$.
However, our formula for the Riemann curvature tensor and the Gauss-Bonnet theorem gives
\begin{align}
\chi^{mn}
&=c_1^n-c_1^m\notag\\
&=2c_1^n-(c_1^n+c_1^m),
\end{align}
which can be an odd integer when the total Chern number of the $n$ and $m$ bands is an odd number.
This shows that there is a topological obstruction to defining a submanifold generated by $\hat{e}^{mn}_a$.

We can derive a similar result for real manifolds ($PT$-symmetric cases with $(PT)^2=1$).
The real projective line ${\rm {\bb R}P}^1\simeq S^1$ is the only compact and connected one-dimensional manifold~\cite{milnor1997topology}.
While $\chi({\rm {\bb R}P}^1)=0$, meaning that $\chi^{mn}=0$, we have in general [See Appendix~\ref{sec:SW}]
\begin{align}
\chi^{mn}\mod 2
&=w_1^n-w_1^m\notag\\
&=w_1^n+w_1^m,
\end{align}
where $w_1=0$ or $1$ is the first Stiefel-Whitney number (Berry phase divided by $\pi$).

We can extend these to the quaternionic manifolds [$(PT)^2=-1$].
In this case, the quaternionic projective line ${\rm {\bb H}P}^1\simeq S^4$ has the Euler characteristic of two.
So, $\chi^{mn}=N_w\chi({\rm {\bb H}P}^1)=2N_w$ is expected.
However, an obstruction can arise due to the relation
\begin{align}
\chi^{mn}
&=p_1^n-p_1^m\notag\\
&=2p_1^n-(p_1^n+p_1^m),
\end{align}
where $p_1$ is the Pontryagin number, when $p_1^n+p_1^m$ is odd.

\subsection{Generalized Gauss-Bonnet theorem in four dimensions}
\label{sec:GB}

The generalized Gauss-Bonnet theorem~\cite{allendoerfer1943gauss} in four dimensions are relevant to the case with twofold degeneracy by $(PT)^2=-1$, in which case we have
\begin{align}
\chi^{mn}
&=\int dk_1dk_2dk_3dk_4\sum_{a_i,b_j}\frac{{\rm sgn}(\wedge d^4k)}{(2\pi)^22^42\sqrt{g}}
\epsilon^{a_1a_2a_3a_4}\epsilon^{b_1b_2b_3b_4}\notag\\
&\times R^{mn}_{a_1a_2b_1b_2}R^{mn}_{a_3a_4b_1b_4}.
\end{align}
We show that the Euler number can be related to the difference of the Pontryagin number of bands $n$ and $m$.

First, we investigate the symmetry constraints.
$PT$ symmetry $PT\ket{u_q}=G_{pq}\ket{u_p}$ imposes $(r^a)_{m_in_j}^*=-(G^{-1}r^aG)_{m_in_j}$.
If we choose a basis where $G_{n_in_j}=(i\sigma_y)_{n_in_j}$ and $G_{m_im_j}=(i\sigma_y)_{m_im_j}$,
\begin{align}
\left(r^a_{m_in_j}\right)
&=
\begin{pmatrix}
r^a_{m_1n_1}&r^a_{m_1n_2}\\
r^a_{m_2n_1}&r^a_{m_2n_2}.
\end{pmatrix}\notag\\
&=-if^a_1\sigma_0+f^a_2\sigma_x+f^a_3\sigma_y+f^a_4\sigma_z,
\end{align}
where
\begin{align}
f^a_1
&=\frac{i}{2}\left(r^a_{m_1n_1}+r^a_{m_2n_2}\right),\notag\\
f^a_2
&=\frac{1}{2}\left(r^a_{m_1n_1}+r^a_{m_2n_2}\right),\notag\\
f^a_3
&=\frac{i}{2}\left(r^a_{m_1n_2}-r^a_{m_2n_1}\right),\notag\\
f^a_4
&=\frac{1}{2}\left(r^a_{m_1n_2}-r^a_{m_2n_1}\right).
\end{align}
For our purpose, it is convenient to introduce the notation
\begin{align}
\left(r^a\right)
&=
\begin{pmatrix}
0&0&r^a_{m_1n_1}&r^a_{m_1n_2}\\
0&0&r^a_{m_2n_1}&r^a_{m_2n_2}\\
r^a_{n_1m_1}&r^a_{n_1m_2}&0&0\\
r^a_{n_2m_1}&r^a_{n_2m_2}&0&0
\end{pmatrix}\notag\\
&=
f_1^a\tau_y\sigma_0
+f_2^a\tau_x\sigma_x
+f_3^a\tau_x\sigma_y
+f_4^a\tau_x\sigma_z\notag\\
&\equiv
f_1^a\Gamma_1
+f_2^a\Gamma_2
+f_3^a\Gamma_3
+f_4^a\Gamma_4
\end{align}
where
$\Gamma_1
=\tau_y\sigma_0,
\Gamma_2
=\tau_x\sigma_x,
\Gamma_3
=\tau_x\sigma_y,
\Gamma_4
=\tau_x\sigma_z,
\Gamma_5
=\tau_z\sigma_0$.
The Gamma matrices satisfy $\Gamma_1\Gamma_2\Gamma_3\Gamma_4\Gamma_5=1$ and thus ${\rm Tr}\left[\Gamma_{i}\Gamma_{j}\Gamma_{k}\Gamma_{l}\Gamma_5\right]=4\epsilon^{ijkl}$.

The determinant of the metric is evaluated to be.
\begin{align}
g
\equiv \det g^{mn}
=\left(4\sum_{a_1,a_2,a_3,a_4}\epsilon^{a_1a_2a_3a_4}f^{a_1}_1f^{a_2}_2f^{a_3}_3f^{a_4}_4\right)^2.
\end{align}

The Riemann curvature tensor can be written simply as
\begin{align}
R^{mn}_{a_1a_2b_1b_2}
&=\frac{1}{2}\sum_{i,j}{\rm Tr}\left[F_{a_1a_2}{\cal F}_{b_1b_2}\right],
\end{align}
where we define
\begin{align}
(F_{a_1a_2})
&=
\begin{pmatrix}
(F_{a_1a_2}^{nm})_{m_im_j}&0\\
0&(F_{a_1a_2}^{mn})_{n_in_j}
\end{pmatrix},\notag\\
({\cal F}_{a_1a_2})
&=
\begin{pmatrix}
({\cal F}_{a_1a_2})_{m_im_j}&0\\
0&({\cal F}_{a_1a_2})_{n_in_j}
\end{pmatrix},
\end{align}
and
\begin{align}
(F_{a_1a_2}^{mn})_{n_in_k}
&=i\sum_j(r^{a_1}_{n_im_j}r^{a_2}_{m_jn_k}-r^{a_2}_{n_im_j}r^{a_1}_{m_jn_k}),\notag\\
(F_{a_1a_2}^{nm})_{m_im_k}
&=i\sum_j(r^{a_1}_{m_in_j}r^{a_2}_{n_jm_k}-r^{a_2}_{m_in_j}r^{a_1}_{n_jm_k}).
\end{align}
Therefore,
\begin{align}
&\epsilon^{a_1a_2a_3a_4}\epsilon^{b_1b_2b_3b_4}R^{mn}_{a_1a_2b_1b_2}R^{mn}_{a_3a_4b_1b_4}\notag\\
&=\frac{1}{4}
\sum_{\mu,\nu,\rho,\sigma=1}^4\left[
\epsilon^{a_1a_2a_3a_4}
(F_{a_1a_2})_{\mu\nu}(F_{a_3a_4})_{\rho\sigma}
\right]\notag\\
&\qquad\times
\left[
\epsilon^{b_1b_2b_3b_4}
({\cal F}_{b_1b_2})_{\mu\nu}({\cal F}_{b_3b_4})_{\rho\sigma}
\right]
\end{align}
The $FF$ term is
\begin{align}
&\sum_{a_1,\hdots,a_4}\epsilon^{a_1a_2a_3a_4}
(F_{a_1a_2})_{\mu\nu}(F_{a_3a_4})_{\rho\sigma}\notag\\
&=-\left(4\sum_{a_1,a_2,a_3,a_4}\epsilon^{a_1a_2a_3a_4}f^{a_1}_1f^{a_2}_2f^{a_3}_3f^{a_4}_4\right)\notag\\
&\qquad\times
\sum_{i,j,k,l}\epsilon^{ijkl}
(\Gamma_i\Gamma_j)_{\mu\nu}
(\Gamma_k\Gamma_l)_{\rho\sigma}.
\end{align}
Now the Gamma matrix part is
\begin{align}
&\sum_{i,j,k,l}\epsilon^{ijkl}
(\Gamma_i\Gamma_j)_{\mu\nu}
(\Gamma_k\Gamma_l)_{\rho\sigma}\notag\\
&=
\frac{3}{2}\left[\delta_{\mu\sigma}(\Gamma_5)_{\rho\nu}
+(\Gamma_5)_{\mu\sigma}\delta_{\rho\nu}
-\frac{1}{2}\delta_{\mu\nu}(\Gamma_5)_{\rho\sigma}
-\frac{1}{2}(\Gamma_5)_{\mu\nu}\delta_{\rho\sigma}
\right]\notag\\
&\quad+\frac{1}{4}\sum_{i,j,k,l}\epsilon^{ijkl}
(\Gamma_i\Gamma_j)_{\mu\nu}
(\Gamma_k\Gamma_l)_{\rho\sigma},
\end{align}
where we apply the Fierz identity twice (see~\ref{sec:Fierz}).
It can be equivalently written as
\begin{align}
&\sum_{i,j,k,l}\epsilon^{ijkl}
(\Gamma_i\Gamma_j)_{\mu\nu}
(\Gamma_k\Gamma_l)_{\rho\sigma}\notag\\
&=
2\left[\delta_{\mu\sigma}(\Gamma_5)_{\rho\nu}
+(\Gamma_5)_{\mu\sigma}\delta_{\rho\nu}
-\frac{1}{2}\delta_{\mu\nu}(\Gamma_5)_{\rho\sigma}
-\frac{1}{2}(\Gamma_5)_{\mu\nu}\delta_{\rho\sigma}
\right].
\end{align}
Using this result, we have
\begin{align}
&\epsilon^{a_1a_2a_3a_4}\epsilon^{b_1b_2b_3b_4}R^{mn}_{a_1a_2b_1b_2}R^{mn}_{a_3a_4b_1b_4}\notag\\
&=-2\det (f^a_i)
\sum_{\mu,\nu,\rho,\sigma=1}^4
\bigg[\delta_{\mu\sigma}(\Gamma_5)_{\rho\nu}
+(\Gamma_5)_{\mu\sigma}\delta_{\rho\nu}\notag\\
&-\frac{1}{2}\delta_{\mu\nu}(\Gamma_5)_{\rho\sigma}
-\frac{1}{2}(\Gamma_5)_{\mu\nu}\delta_{\rho\sigma}
\bigg]\left[
\epsilon^{b_1b_2b_3b_4}
({\cal F}_{b_1b_2})_{\mu\nu}({\cal F}_{b_3b_4})_{\rho\sigma}
\right]\notag\\
&=4\det (f^a_i)
\epsilon^{b_1b_2b_3b_4}
{\rm Tr}
\left(
-\Gamma_5{\cal F}_{b_1b_2}{\cal F}_{b_3b_4}
\right).
\end{align}
By combining this expression and and the expression of the determinant of the metric and using ${\rm sgn}[\det (f^a_i)]={\rm sgn}(\wedge d^4k)$, we finally obtain
\begin{align}
\chi^{mn}
&=\int dk_1dk_2dk_3dk_4\sum_{b_j}\frac{2^2}{(2\pi)^22^42}
\epsilon^{b_1b_2b_3b_4}\notag\\
&\quad\times{\rm Tr}
\left(-
\Gamma_5{\cal F}_{b_1b_2}{\cal F}_{b_3b_4}
\right)\notag\\
&=\int \frac{d^4k}{32\pi^2}\sum_{b_j}
\epsilon^{b_1b_2b_3b_4}\notag\\
&\quad\times\sum_i\left[
({\cal F}_{b_1b_2}{\cal F}_{b_3b_4})_{n_in_i}
-({\cal F}_{b_1b_2}{\cal F}_{b_3b_4})_{m_im_i}
\right]
\notag\\
&=p_1^n-p_1^m.
\end{align}

\subsection{First Stiefel-Whitney number in one dimension}
\label{sec:SW}

Here we consider $PT$-symmetric systems with $(PT)^2=1$.
Let us choose a direction $a$ such that $r^a_{mn}$ is nonvanishing along a one-dimensional curve in the Brillouin zone, then the first Stiefel-Whitney number on the curve is given by
\begin{align}
w_1^{mn}
&=\frac{1}{\pi}\oint d{\bf k} \cdot i\d_{\bf k} \log r^a_{mn}.
\end{align}
This formula is calculated in a complex smooth and periodic gauge.

The first Stiefel-Whitney number $w_1^{mn}$ measures whether $r^a_{mn}$ in a real gauge (where eigenstates $\ket{n}$ and $\ket{m}$ are real such that $r^a_{mn}$ is also real) reverses its direction as it goes around the one-dimensional closed curve.
To see this, since $r^a_{mn}$ is real-valued in a real gauge, it has zero phase winding.
It means that, if it had a nonzero phase winding $w_1^{mn}\ne 0$ in the smooth gauge, the gauge transformation $\{\ket{n},\ket{m}\}\rightarrow \{e^{i\phi_{n}}\ket{n},e^{i\phi_m}\ket{m}\}$ from the complex gauge to real gauge (i.e., $r^a_{mn}\rightarrow e^{-i(\phi_n-\phi_m)}r^a_{mn}$) should satisfy $\frac{1}{\pi}\oint d{\bf k} \cdot \d_{\bf k} (\phi_n-\phi_m)=w_1^{mn}$ to cancel the nontrivial phase winding.
However, this kind of gauge transformation induces a nontrivial boundary condition on the real-valued tangent vector, $r^a_{mn}(2\pi)=e^{i[\phi_{nm}(2\pi)-\phi_{nm}(0)]}r^a_{mn}(0)=e^{i\pi w_1^{mn}}r^a_{mn}(0)$ in real gauge, since $r^a_{mn}(2\pi)=r^a_{mn}(0)$ in the initial complex smooth and periodic gauge.
This means that $r^a_{mn}$ in the $w_1^{mn}=1$ case do not form a tangent vector of a real manifold, because it is not a smooth and single-valued vector on a circle.

Let us now show that $w_1^{mn}$ can be related to topological invariants of bands $n$ and $m$ as in the case of the Euler characteristic of complex manifolds.
That is, we show that
\begin{align}
\label{eq:SW-relation}
w_1^{mn}=w_1^n-w_1^m,
\end{align}
where $w_1^n$ and $w_1^m$ are the first Stiefel-Whitney numbers of bands $n$ and $m$.
Since $w_1^n$ is identical to the Berry phase of band $n$ divided by $\pi$~\cite{ahn2018band,ahn2019stiefel}, i.e.,
\begin{align}
w_1^n=\frac{1}{\pi}\oint d{\bf k}\cdot {\bf \cal A},
\end{align}
Eq.~\eqref{eq:SW-relation} follows from that the real part of the following shift vector $R^{c,a}=i(r^a_{mn})^{-1}\d_cr^a_{mn}+({\cal A}^c_{mm}-{\cal A}^c_{nn})$ vanishes in $PT$-symmetric systems.

We can show ${\rm Re}R^{c,a}=0$ as follows.
The $PT$ symmetry constraints $PT\ket{n}=e^{i\theta_{n}}\ket{n}$ and $PT\ket{m}=e^{i\theta_{m}}\ket{m}$ lead to $r_{mn}=i|r_{mn}|e^{i(\theta_m-\theta_n)/2}$, ${\cal A}^a_{mm}=\frac{1}{2}\d_a\theta_m$, ${\cal A}^a_{nn}=\frac{1}{2}\d_a\theta_n$ for $m\ne n$.
Therefore, we have $i(r^a_{mn})^{-1}\d_cr^a_{mn}=-\frac{1}{2}(\d_c\theta_m-\d_c\theta_n)+i|r^a_{mn}|^{-1}\d_c|r^a_{mn}|$.

\subsection{Vanishing of the torsion tensor on generalized flag manifolds}

Let us define
\begin{align}
\hat{e}_a
&=\sum_{m>n}\ket{u_m}(r^a_{mn}-{\cal A}^a_{mn})\bra{u_n},
\end{align}
which is the complexification of the Maurer-Cartan form $\hat{\xi}_a=\hat{e}_a+\hat{e}^{\dagger}_a$ of a generalized complex flag manifold
\begin{align}
{\rm Fl}_{\bb C}(N_1,\hdots,N_k)
=\frac{U(N)}{{\rm U}(1)\times\hdots\times {\rm U}(N_k)},
\end{align}
and ${\cal A}^a$ is the ${\rm U}(1)\times\hdots\times {\rm U}(N_k)$ Berry connection.
The Hermitian metric is defined by
\begin{align}
Q_{ba}=(\hat{e}_b,\hat{e}_a)={\rm Tr}[\hat{e}^{\dagger}_b,\hat{e}_a],
\end{align}
and the Hermitian connection is defined by
\begin{align}
C_{bca}=(\hat{e}_b,\nabla_c\hat{e}_a)=(\hat{e}_b,\d_c\hat{e}_a).
\end{align}
The torsion tensor is then defined by
\begin{align}
T_{bca}
&=C_{bca}-C_{bac}-i(\hat{e}_b,[\hat{e}_c,\hat{e}_a]).
\end{align}
To show that torsion is zero. let us calculate the commutator.
\begin{align}
&(\d_c\hat{e}_a-\d_a\hat{e}_c)_{mn}\notag\\
&=
\d_c(r^a_{mn}-{\cal A}^a_{mn})\notag\\
&\qquad
-i\sum_p\left[{\cal A}^c_{mp}(r^a_{pn}-{\cal A}^a_{pn})
-(r^a_{mp}-{\cal A}^a_{mp}){\cal A}^c_{pn}\right]
-(c\leftrightarrow a)\notag\\
&=
i\sum_p(r^c_{mp}r^a_{pn}-r^a_{mp}r^c_{pn})
-{\cal F}^{ca}_{mn}\notag\\
&\qquad
-i\sum_p\bigg[{\cal A}^c_{mp}r^a_{pn}+r^c_{mp}{\cal A}^a_{pn}
-r^a_{mp}{\cal A}^c_{pn}-{\cal A}^a_{mp}r^c_{pn}\notag\\
&\qquad
+({\cal A}^c_{mp}{\cal A}^a_{pn}-{\cal A}^a_{mp}{\cal A}^c_{pn})
\bigg]
\notag\\
&=
-{\cal F}^{ca}_{mn}
+i\sum_p
\bigg[
(r^c_{mp}-{\cal A}^c_{mp})(r^a_{pn}-{\cal A}^a_{pn})\notag\\
&\qquad-(r^a_{mp}-{\cal A}^a_{mp})(r^c_{pn}-{\cal A}^c_{pn})
\bigg]
\notag\\
&=
-{\cal F}^{ca}_{mn}
+i[\hat{e}_c,\hat{e}_a]_{mn}.
\end{align}
Because $(\hat{e}_b,{\cal F}^{ca})=0$, we have $C_{bca}-C_{bac}=(\hat{e}_b,\d_c\hat{e}_a-\d_c\hat{e}_a)=i(\hat{e}_b,[\hat{e}_c,\hat{e}_a])$, and thus
\begin{align}
T_{bca}
&=0.
\end{align}
One can also show that $T_{bca}=0$ on generalized real and quaternionic flag manifolds.

\appendix

\section{Fierz identities}
\label{sec:Fierz}

Here we follow Ref.~\cite{nieves2004generalized}.
Let us introduce the following notation $\Gamma^{\mu}_A$ for the 16 generators of the $4\times 4$ matrices.
Here, $A$ indicates the representation under the action of $O(4)$, and $\mu$ indicates the components in it.
\begin{align}
\Gamma^{\mu=1}_S
&=1,\notag\\
\Gamma^{\mu=1,\hdots,4}_V
&=\Gamma_{\mu},\notag\\
\Gamma^{\mu=1,\hdots,6}_T
&=-i\Gamma_i\Gamma_j,\notag\\
\Gamma^{\mu=1,\hdots,4}_A
&=i\Gamma_5\Gamma_\mu,\notag\\
\Gamma^{\mu=1}_P
&=\Gamma_5,
\end{align}
where $\{\Gamma_i,\Gamma_j\}=\delta_{ij}$ for $i,j=1,\hdots,5$.

The Fierz identity follows from the completeness of these basis matrices.
That is, for any $4\times 4$ matrix $M$, we can express it as
\begin{align}
M=\sum_{\mu,A}m^{\mu}_A\Gamma^{\mu}_A
\end{align}
for some $m^{\mu}_A$.
It's value given by
\begin{align}
m^{\mu}_A
=\frac{1}{4}{\rm Tr}(\Gamma^{\mu}_AM),
\end{align}
where we use (no Einstein summation convention below)
\begin{align}
{\rm Tr}(\Gamma^{\mu}_A\Gamma^{\nu}_B)
&=4\delta_{\mu\nu}\delta_{AB}.
\end{align}
Rewriting the above completeness relation,
\begin{align}
M_{ab}
&=\sum_{\mu,A}\frac{1}{4}{\rm Tr}(\Gamma^{\mu}_AM)(\Gamma^{\mu}_A)_{ab}\notag\\
&=\frac{1}{4}\sum_{\mu,A}\sum_{c,d}(\Gamma^{\mu}_A)_{cd}M_{dc}(\Gamma^{\mu}_A)_{ab}\notag\\
&=\sum_{c,d}\left[\frac{1}{4}\sum_{\mu,A}\Gamma^{\mu}_A\Gamma^{\mu}_A\right]_{ab;cd}M_{dc},
\end{align}
one can find
\begin{align}
\frac{1}{4}\sum_{\mu,A}(\Gamma^{\mu}_A)_{ab}(\Gamma^{\mu}_A)_{cd}
=\delta_{ad}\delta_{bc}.
\end{align}

Let us multiply $(\Gamma^{\nu}_B)_{b'b}(\Gamma^{\nu}_B)_{d'd}$ and sum over $\nu,a,c$.
\begin{align}
\sum_{\nu}(\Gamma^{\nu}_B)_{ad'}(\Gamma^{\nu}_B)_{cb'}
&=\frac{1}{4}\sum_{\mu,\nu,A}(\Gamma^{\mu}_A\Gamma^{\nu}_B)_{ab'}(\Gamma^{\mu}_A\Gamma^{\nu}_B)_{cd'}\notag\\
&=\sum_{\rho,\sigma,C}(\Gamma^{\rho}_C)_{ab'}(\Gamma^{\sigma}_C)_{cd'}\left(\frac{1}{4}\sum_{\mu,\nu,A}f_{CAB}^{\rho\mu\nu}f_{CAB}^{\sigma\mu\nu}\right)\notag\\
&=\sum_{\rho,C}(\Gamma^{\rho}_C)_{ab'}(\Gamma^{\rho}_C)_{cd'}K_{BC},
\end{align}
where we use that $\sum_{\mu,\nu,A}f_{CAB}^{\rho\mu\nu}f_{CAB}^{\sigma\mu\nu}\propto \delta_{\rho\sigma}$ because it should be $O(4)$-invariant.
The value of $K$ can be calculated by contracting $a$ with $b'$ and $c$ with $d'$:
\begin{align}
{\rm Tr}(\Gamma^{\nu}_B\Gamma^{\sigma}_D\Gamma^{\nu}_B\Gamma^{\sigma}_D)
=\sum_{\rho,C}\left[{\rm Tr}(\Gamma^{\rho}_C\Gamma^{\sigma}_D)\right]^2K_{BC}
=16K_{BD}
\end{align}
So,
\begin{align}
K_{AB}
=\frac{1}{16}\sum_{\mu,\nu}{\rm Tr}(\Gamma^{\mu}_A\Gamma^{\nu}_B\Gamma^{\mu}_A\Gamma^{\nu}_B).
\end{align}

\begin{align}
\left(K_{AB}\right)
=\frac{1}{4}
\begin{pmatrix}
1&1&1&1&1\\
4&-2&0&2&-4\\
6&0&-2&0&6\\
4&2&0&-2&-4\\
1&-1&1&-1&1
\end{pmatrix}.
\end{align}


%

\end{document}